# PERSONAL HISTORY OF MY ENGAGEMENT WITH CUPRATE SUPERCONDUCTIVITY, 1986-2010

PHILIP W ANDERSON

ABSTRACT

Six months ago I was asked to write a personal history of my engagement with the high-$T_c$ problem of the cuprate superconductors, in rather informal and autobiographical style. As the work proceeded I realized that it was impossible and would have been dishonest to separate out my rather amusing but seminal early fumblings from the complete restructuring of the problem which I have achieved during the past decade. But the result became considerably too long, by over half, for its intended recipient. The assignment had left me with no obligation to deal with all the fascinating but irrelevant phenomenology which I had more or less instinctively ignored on my way, but that feature also fails to endear the article to any conceivable editorial board containing knowledgeable experts on the subject. Also, their purpose was for it to serve as an "introduction to the more technical debates", but its message is that almost all of these are not relevant. They are not, on the whole, focused on achieving understanding of the crucial experimental anomalies, many, if not most, of which are now understood. The key to the problem is a new method of dealing with the constrained Hilbert space which follows from the necessity of Gutzwiller projection.

1.  INTRODUCING PWA

I must introduce myself.  I'm the scion of a midwestern academic family, sent off to Harvard in 1940 at 16 to take one of Harvard's extraordinarily liberal National Scholarships which presaged the selective Ivy League of today.  After an accelerated 3 years, and a bit of war work, I wanted to experience Harvard as a more mature individual, so went back there to grad school.  I was fortunate enough to study theoretical physics under a marvelous man named John van Vleck, (hereafter "Van") who gave me a good problem, the right place to start on it, and freedom.  The prejudices of my somewhat impecunious academic family and the fact of an early marriage sent me, after my degree in '49, to industry--Bell Labs--rather than into the academic treadmill of the time.  This fortunate choice landed me at the core of the burgeoning field of solid state physics, learning at the feet of its founders--Bardeen, Kittel, Shockley, Wannier, Herring and others.  It was 35 years before I left Bell, coincidentally with the fateful, idiotic decision of our government to smash into bits the greatest technological institution the world has ever seen.  But of those 35 years I had spent large parts of about 20 on leave at various universities: Tokyo, Cambridge, and Princeton.  After leaving Bell, I became a full-time professor at the last of these, where I had at first been rather a lonely outlier among the astro-, mathematical, and elementary particle physicists,  but by the time of this story I had begun to have first-rate colleagues, as you will learn. In 1977, I was--honestly--surprized to be chosen for the Nobel Prize along with my beloved Van and another patron and senior colleague, Sir Neville Mott of Cambridge.  The main motivation for my prize was my idea of quantum "localization", which has little to do with our subject matter here; but a subsidiary theme was the idea of "superexchange" in magnetic insulators, and its relation to the Mott insulator, which will turn out to be crucial as we shall see. The Prize is a fantastic ego-booster, of course.  The Bell Labs even made a little movie about me, and everyone seemed to want my opinion on all kinds of things--though not, of course, to listen to it on such matters as Star Wars (aka SDI).  I spent time on the council of the National Academy and on the executive committee of the American Physical Society, and was offered a big job as director of the Institute of Theoretical Physics (ITP) at Santa Barbara, agonizing over that for several months but not accepting it. And so I suppose I came to assume I had a level of infallibility, which assumption was both unrealistic and unwise, given my obvious weaknesses: a tendency to jump to conclusions, a fairly short temper, a certain inarticulateness; and for any administrative job, an allergy to balance sheets.

During the decade after the Prize science worked very well for me. Localization became a 'growth industry", as I claimed in one of my talks, and a worldwide network of friendly collaborators, meeting annually at the Aspen Center of Physics, and also involving my growing group at Princeton, formed around it. Names we will encounter later are El Abrahams, Hide Fukuyama, Rama Ramakrishnan, and Patrick Lee. Another field which I jointly originated, spin glasses, became fashionable and useful, and led me into contact with the fascinating world of complexity theory and to ties to the Santa Fe Institute. These were only two of the remarkable developments on which we worked during that period. It seemed as if Condensed Matter, as we had renamed solid state physics, was going to continue to spawn exciting new ideas and phenomena indefinitely, entirely outside of the humdrum technological world from which it had arisen. My combative spirit was aroused only by the difficulty of getting great work adequately supported, and especially of attracting younger colleagues and persuading my department to hold on to them.

2. JANUARY 1987: THE CUPRATES CHANGE MY LIFE, TOO

It is not widely known that John Bardeen produced at least three wrong theories of superconductivity before BCS, the one which got it right. Two of his previous attempts were published, one, in 1951, with great fanfare. The difference between John and the many other brilliant physicists who attempted theories of superconductivity (among them Einstein, Feynman and Heisenberg ) was that he was willing to admit that he had been wrong, go back to the beginning and start over.

This at least I have in common with John; as I examine the 24 years that have ensued since the Mueller discovery, and look around the continual proliferation of "schools" emphasizing one or another special set of data or theoretical gimmick as the "key" to the high-$T_c$ cuprate superconductors, I do not find many colleagues and competitors who have given up on their first insight, gone back to the beginning and started over. This gives me an enormous advantage, and I'm not above using it. The other advantage of this mode of working is that among the wreckage of a dismantled previous theory there may remain a few gems of truth, so you are never quite starting over from scratch.

One of the strange things about this personal history is that my first stab at it was far closer than some of its successors; my random walk was sometimes in the wrong direction. This first attempt dates to December 1986 and January 1987.

In those times I had not completely been weaned from Bell Labs, so had spent there some days during Christmas vacation, as did many other Bell alumni. At Bell I met my close friend Ted Geballe, past supervisor and collaborator of the charismatic Bernd Matthias, who is rightly considered the father of the field of superconducting materials. Ted, it turned out, had just returned from the Materials Research Society meeting in Boston, the first week of December, where a highly reliable Japanese researcher, Kitazawa, had surprised everyone by confirming unequivocally the results on a new superconductor, $(La,Ba)_2CuO_4$, with a transition temperature over 30 degrees, which had been published obscurely six months earlier by Georg Bednorz and Alex Muller. Ted, although a very down-to-earth empiricist, who shares to an extent Bernd's disdain for theories not his own, is somewhat optimistic on the subject of high-$T_c$ superconductors, and has been known to fall for an occasional USO (unidentified superconducting object); but he convinced me this one was real. Paul Chu of Houston also confirmed the new results at this meeting.

Kitazawa had continued his itinerary to give a talk at Bell Labs, from which talk a ceramicist named Bob Cava slipped out early to start up his own furnaces--which, before the end of the year, had already bettered Muller by nearly ten degrees by substituting Sr for Ba. Already there was no doubt that something extraordinary was really, reproducibly happening. The story of the frantic race among the materials makers touched off by the MRS meeting (or in Paul Chu's and Tanaka's cases, by their reading the original Swiss paper) has been very well told, so I won't dwell on it. The key happening was that, by a hair, M K Wu at Alabama, stimulated by Paul Chu, was the first of 4 or 5 independent groups to observe the unlikely fact that yet another compound (soon identified as $YBa_2Cu_3O_7$) of similar composition and structure, extended the range of $T_c$ to above 90 degrees K.

A second fact is that these materials are easy to make, and very reliable--no problem with reproducibility here, as the dozens of experimental reports at "Woodstock" (N Y, March '87) made clear. The series of dramatic announcements caused an entirely disproportionate media furor.

I think it was Ted who told me about the structure, and that when I then was explaining it to Baskaran I began to see the picture. Baskaran will play a big role in what follows. He had come to me from the ICTP(International Centre of Theoretical Physics) in Trieste, with the warm recommendation of my old student from Cambridge, Erio Tosatti. Baskaran is brilliant and quick off the mark. Perhaps it is a Tamil trait that he speaks loudly and with assurance--certainly he shares it with others I know. But perhaps a component of the unfolding contretemps is the fact that his conclusion-jumping resonates with my own. Baskaran was to be enormously supportive and helpful, but occasionally he was so whether I was right or wrong.

Ted described the superconductor to me as "a kind of perovskite". The perovskite structure may be the very first crystal structure I became intimate with, because in 1949, before I had really learned solid state physics, I had been willy-nilly plunged into thinking about the ferroelectric $BaTiO_3$, a canonical perovskite, by Bill Shockley. What I realized was that the structure was two layers of composition LaO, and then a layer of $CuO_2$. If La, as it always and stably is, is a 3+ ion, and O as always behaves like $O^{--}$, that makes $Cu^{++}$ for charge balance. If, as Bednorz and Muller had done, one replaces 10% of the La with $Ba^{++}$--again a very stable valence--then what seems likely is that some of the Cu will become $Cu^{+++}$, a possible, if not a very common, valence for Cu. Right from the start, I assumed that it was these positively charged "holes" in the $Cu^{++}$ $d$-shell which were the metallic carriers--as had the inventors, but not everyone else. It was known that pure $La_2CuO_4$ was an insulator.

What I realized--and what undoubtedly had a lot to do with the unforgivable arrogance that I was to exhibit in the next years-- was that my entire career for nearly 40 years had uniquely prepared me to deal with these materials. That I started out working with perovskites was maybe a little coincidence; but one of the major themes of my work for the first decade was the key concepts of Mott insulator and superexchange; and my own professor in the 30's had invented the crucial idea of "Jahn-Teller" distortions --too modest to name them after himself--which Muller talked about and which really do play a role. I didn't myself invent antiferromagnetism--which is a big part of the story-- but I wrote many of the early important papers on it. AND I'd spent the rest of that early decade or so working on superconductivity!

First, "$d$-shells". The $d$ electron states are concentrated in the interior of the atom, and there are five of each spin. The first few that go in aren't all that isolated and participate in bonding pretty well--hence the great strength and high melting points of Cr, V, Fe--even more so W and Nb, in the next row. But as bonding $d$-states fill up, the last few tend to be somewhat "antibonding" and isolated, and their spins are free to exhibit magnetism. So the canonical magnetic metals are Fe, Co, Ni, and the magnetic ions $Mn^{++}$, $Ni^{++}$, $Fe^{+++}$, $Cu^{++}$. (These are also the colored ones--ochre, blue vitriol.) Let's focus on Cu. With two missing electrons, one of them came from the metallic s shell but one has to come out of the $d$ shell, so there's one odd spin which we ascribe to a "$d$ hole". The $d$ hole can be in any one of the five $d$-shell orbitals, which in the free ion in spherical symmetry all have the same energy--how to choose? But in crystals they aren't spherically symmetric; in a perovskite it has an octahedron of $O^{--}$ ions around it, so to first approximation it's cubic in symmetry. This divides the states up into two groups, the ones with symmetry like $xy$, $yz$, $zx$ of which there are three; and the two with symmetry $x^2$-$y^2$ and $z^2$. These have lobes which point toward the $O^{--}$ ions and therefore are more unhappy--higher in energy than the others--so the hole will be in one of them. This splitting into two sets of $d$-levels was initially ascribed to "crystal fields"; that is, the $d$ electrons were seen as feeling the repulsion from the negatively charged oxygen ions. In the thirties Van pointed out a more realistic and quantitative way to think about it, when he introduced the concept of a "semicovalent bond". The idea was that the oxygen "$p$" electrons which point toward the copper (say) would partially mix with the $d$ electrons of the appropriate shape, lowering their energy thereby and forming a weakish bond.  Then when we try to put an electron on the Cu its energy is raised because it has to form an unsatisfactory odd linear combination --an "antibonding orbital"--with the oxygen.  Its amplitude must go through zero just where the combined potential of the two atoms is most attractive.  After WWII these ideas were elaborated on by chemists and there were excellent books discussing them.  Bob Shulman and his collaborators (of whom I was one) spent a lot of time in the late'50's checking them out by means of the "transferred hyperfine structure", the magnetic fields caused at the "ligand" nucleus by the admixture of the $p$ level into the magnetic $d$ one.  $Cu^{++}$ plays out this scenario in a particularly acute form.  This ion has only one $d$ hole--only one electron missing from the full shell.  So it may choose either "$d_{x2-y2}$" or "$d_{z2}$" symmetry.  It normally chooses the former, thus allowing there to be one unpaired electron bonding with the square of oxygen atoms around it.  Thus it draws them in towards it, and correspondingly pushes away the "apical oxygens" at the top and bottom of its octahedron.  This and its relatives Van modestly called the "Jahn-Teller effect", although those two gentlemen hadn't had any idea of applying their idea about structures of molecules to this kind of system.  The presence of this Jahn-Teller distortion is unequivocal evidence that there is a large splitting between the "bonding", predominantly oxygen, $p$ orbital and the "antibonding" level where the $d$ holes are to be found.

The Bell group was able to show me quantitatively the structure of the superconducting material, and it was clear that this distortion was very strong in the La compound--that the $CuO_2$ layer was definitely quite tightly pulled in, with the apical oxygens of the LaO layers considerably farther from Cu, so that the octahedron of oxygens was quite pointy.  It was, in fact, either the possibility or the actuality--I don't know which—of this J-T effect which drew Muller's attention to this compound originally. I could speculate that he thought that the distortion due to two holes would be twice as big as for one, and the energy gained more than twice—it might even more than make up for "$U$", the extra energy due to repulsion of the two positive charges.  Therefore the Cu could be what I once called a "two-electron center" or "negative $U$", preferring to be $Cu^{++} Cu^{+++}$ rather than two $Cu^{++}$'s.  Such objects actually occur in some systems, but they don't make superconductors, even though they amount to electrons strongly attracted in pairs.  This is because the ions don't move quantum-mechanically--the lattice distortion must move along with the ion to keep it stable, so it behaves as though the electron were very heavy.  In the cuprates the fact that the effective $U$ is, on the contrary, positive and large was well-known, because the pure compound $La_2CuO_4$ is a "Mott Insulator".  I suspect this was told me by Baskaran, since the references I used in January '87 are from India in '85, and I wouldn't have known them otherwise.  The Mott insulators are an enormous class of compounds of transition metals, rare earths and the like which are insulators not because of the standard Wilson argument that their energy bands are all full of pairs of electrons, but because the electron repulsion $U$ is so strongly positive that all of the ions must have the same charges even though that leaves magnetic, open shells.  A " Wilson" (conventional) insulator cannot be magnetic, a Mott insulator usually is.  But the J-T distortion does play an important and interesting role.  What it does is to separate the two energy levels of the $d$ electrons by as much as a volt, so that the mobile states are entirely of $d_{x^2-y^2}$ symmetry:  there is only one relevant state per Cu atom.  This state may be empty or singly occupied--$Cu^{+++}$ or $Cu^{++}$-- but that is the totality of the options there are.  It happens that due to pure coincidence, the amount of Jahn-Teller distortion that makes either of these states happy is about the same, so that the $d$ holes are reasonably mobile, not carrying a great deal of distortion along with them.  On the other hand, a $Cu^+$ is a spherically symmetrical electronic state and does not want or need the distortion--it may not be a mobile carrier at all, because the structure around it wants to be so different.  (Even at this late date, the "electron-doped" compounds which are presumed to have such carriers are impenetrably complex, though showing some lower-temperature superconductivity.)

I began to realize that we were being presented with the perfect exemplar for a mathematical model of great historical importance: the Hubbard Model.  John Hubbard was a theoretical physicist at the Harwell Lab near Oxford--the site of a neutron reactor used to study magnetic materials.  He invented the model as the simplest possible description of a magnetic material which seemed to contain all the necessary physics--in particular, it can describe a Mott insulator when there is exactly one electron per site.

He presumes only one energy band, which means there is only one relevant wave function per site, which may of course be occupied by either an up- or down-spin electron or both. The only complication is that he includes a (presumably) rather strong repulsive interaction $U$ when two electrons are on the same site--otherwise the electrons may hop freely from site to site. Though simple, this model remains one of the classic unsolved problems of physics—it has spawned many thousands of papers, but only in very recent times have we even begun to figure out the right questions to ask of it. To have a Mott insulator which not only fit the idealized Hubbard picture to a T, but could be doped away from one electron per site, seemed more good fortune than one could possibly expect. Given that so far the high $T_c$ of cuprates was an utter mystery, and that they turned out to be a nearly perfect instance of another nearly equal mystery, the Hubbard model in its simplest manifestation, I could hardly fail to feel that the two mysteries were related.

At the end of December much of the community which had been galvanized by the cuprate news pulled up stakes and gathered in Bangalore, India. The meeting had been called to discuss a related topic, the so-called "heavy electron" or "mixed valence" phenomenon in (mostly) rare earth compounds, which was a source of some exotic superconductors. But the talk in the bars was all about the new compounds. I left home for Bangalore excited about the cuprates, though not quite in the hyper state I returned in. For one thing, I had no idea that my reasoning was not manifest to everyone who gave any thought to the problem. Certainly my old friend Maurice Rice, with whom Bill Brinkman and I had suffered in the 70's through the attempt to prove that our Bell experimentalists were seeing a Mott transition, would see it as I did. Maurice will turn out to be one of the few heroes of this story, so I should introduce him. He's a tall, dark-haired Irishman, whom I had known since he had been in graduate school in Cambridge in 1961. Eventually, Maurice had moved to the ETH in Zurich (he had married a Swiss wife in California) where he was typically well funded and supplied with research associates. I was to use him on several occasions to rescue struggling students or postdocs; he remained an isolated island of sanity in an almost uniformly bleak European superconductor community. But, actually, he hadn't thought much about the problem. It was not informal discussions but his actual conference paper which triggered the next steps in my reasoning. They begin with the one-dimensional antiferromagnetic chain of magnetic atoms--we'll often call them "spins" and even specialize to the case of spin = ½. Among the other wondrous features of the cuprates is that only spin ½ plays a role. A chain of atoms with positive exchange integral (the energy is written

$$H = -J \sum_{\langle i,j \rangle} \mathbf{S}_i \cdot \mathbf{S}_j \qquad [1]$$

and J is called the exchange integral) has a ferromagnetic ground state, with all spins parallel, and Bloch showed that the excited states had one reversed spin moving as a wave. But the antiferromagnetic case, with *J*<0, is much more complicated. The very difficult problem of exactly what does happen was, amazingly, solved already by Hans Bethe in 1932, in a *tour de force* of mathematical physics which has been a model for many further developments. A main thing which he showed was that the ground state is NOT antiferromagnetically polarized but has no simple kind of "long-range order" at all. Quantum fluctuations of the spins are so large in one dimension that they are unstable and destroy long-range antiferromagnetic "Neel" order.

What Maurice's paper showed was that one could achieve very nearly--but not exactly--as good a solution as Bethe's with a ridiculously simple wave function, constructed as follows: Take one electron per site, but instead of fixing them down at local sites, put them all in the free-particle, plane-wave states that they would occupy if they had no interactions at all with each other, but instead just a hopping energy *t* between the nearest neighbor sites. They would then exactly half-fill the energy band with a Fermi sea of uncorrelated electrons. Now carry out the following procedure, which is called "Gutzwiller projection": The sites are, for free electrons, randomly occupied ½ of the time by up electrons and ½ of the time by down ones, so the probability that any given site is empty is ¼, and the probability that it's doubly full is also ¼. Let us throw away every part of the wave function for which any site is either empty or doubly full, so that on every site there is exactly one electron with spin up or down. You realize that we have thrown away almost all of the wave function, leaving only (½)$^N$ of it; but what we have left has a spin on every site and thus will do as an approximate wave function for the Heisenberg antiferromagnetic chain. What

Maurice had done was to do careful numerical calculations of how well this approximated the energy of Bethe's solution. The answer was, very well indeed, thank you--it was within tenths of a percent. There seems to be, in this free electron wave function, precisely the right degree of spin-spin correlation. I was intrigued by this and, in the course of the next day or so, began to have an inkling why it was so. I realized that, after Gutzwiller projection, this wave function is just the same as that of a superposition of zero-momentum "BCS pairs" such as Bob Schrieffer wrote down to describe ordinary superconductors. The pairing function is rather extreme--one makes all the pairs with momentum vectors **k** inside the Fermi surface have positive sign, and all the ones outside have the opposite, negative sign. As with the ordinary type of BCS function, this has the effect of binding the electrons together in singlet pairs. And one leaves out the kinetic energy entirely--every pair is treated as though it were exactly at the Fermi energy. Formally, the function is the product of factors $(u_\mathbf{k} + v_\mathbf{k} c^\dagger_{\mathbf{k}\uparrow} c^\dagger_{-\mathbf{k}\downarrow})$, one for each **k** value in the zone. $v_\mathbf{k}/u_\mathbf{k}$ is $\pm 1$, depending on whether **k** is inside or outside of Maurice's Fermi surface. From this representation of the wave function springs the core insight which has been the most seductive and the most controversial in all the long years that we have all spent arguing about the subject.

To explain the origins of this we have to go back 15 years to Cambridge. Stimulated by a couple of experimental papers from Japan, I had returned in my thoughts to the old question of the contrast between the disordered Bethe solution of the antiferromagnetic linear chain, and the well-ordered antiferromagnetic "Neel" solution which applied, for sure, in most three-dimensional lattices. The methods I had used to prove antiferromagnetism barely converged in two dimensions, and would be particularly problematical for the triangular lattice which is strongly "frustrated". Therefore I wondered in print[i] whether there could be a Bethe-like state in this case, since the experimental data suggested that there was no order even near $T = 0$. I called this a "resonating valence bond" state, RVB for short, borrowing the term from Linus Pauling who used it to describe metals in a way I didn't think was very useful. In a gesture which might be called ironic but admiring I sent the paper off to a Festschrift for Pauling. Later on, a Hungarian postdoc, Patrick Fazekas, and I elaborated on it somewhat, but we remained mystified about the nature of this liquid-like state; and the experimental basis did not seem to work out either, so it was almost forgotten, even by me. But as a puzzling theoretical question--what would be the properties of such a state, if anyone found it? Does it have any kind of order, or any special responses?—it remained in the back of my mind. The light-bulb which appeared in my mind that evening in Bangalore was that this Gutzwiller projection technique allowed me to write down an explicit wave-function for an RVB – a wave-function which had been shown to be very accurate for the one case we had precise information about, the 1-D insulating chain, but which was generalizable to any lattice of any dimension of our choice. It was seductive to suppose that (a) the actual ground state of even the two-dimensional square lattice was an RVB; and that (b) somehow, as one added holes, which would make the material into a metal, the projected BCS state would become a superconductor.

The idea of the superexchange mechanism had been generalized to metals in a series of papers mostly concerned with the mixed valence problem. The result is something called the *t-J* Hamiltonian, which is a physical idea which had grown up as folk knowledge without it being possible to assign its invention to a particular person. The formalism owes a lot to W Kohn's theory of insulators.[ii] I will give it in a modern version which can be found in a paper of 1986 by Maurice Rice and coworkers.[iii]

If we dope a Mott insulator with holes, there are still no sites with pairs of electrons on them, but there are holes, i.e. empty states which are free to move. We still have to allow for the process of virtual hopping of an electron onto an occupied site. This is done by a "perturbative canonical transformation" which eliminates all matrix elements which do such hops, but leaves behind a superexchange term like Eq. 1 above. We must also allow the kinetic energy to exist, but it must be the projected kinetic energy, in which the electron is not allowed to hop onto a wrong site. This is the "*t*" term of the *t-J* Hamiltonian. The worst mistake is to assume that it will do no harm to forget that it's projected, and this is sometimes done, even by very famous physicists. The *t-J* Hamiltonian, then, is:

$$H_{t-J} = P \sum_{i<j,\sigma} t_{ij} c_{i,\sigma}{}^* c_{j,\sigma} P + \sum_{i,j} J_{ij} S_i \cdot S_j \ . \qquad [2]$$

Here *P* is the full Gutzwiller projector,

$$P = \prod_i (1 - n_{i,\downarrow} n_{i,\uparrow}),  \qquad [3]$$

which eliminates all doubly-occupied states. There is no need to apply it to the *J* term because that does not change site occupancies: it is already projective.

Now, finally, back to the new cuprates. The pure substance $La_2CuO_4$, as I said, is a Mott insulator--and so far, it had not been seen to have an antiferromagnetic structure. So I proposed that it was actually an RVB, described pretty well by my projected BCS pairing function. And then, said I, it stands to reason that when you dope it with a few carriers they will partake of that pairing and be superconducting. The only problem, at first, is that the pairing energy of such an RVB is approximately *J*, so the superconducting $T_c$ would be large--already, it was possible to guess that *J* was extraordinarily big, over 1,000 degrees Kelvin. The solution to that was offered by Baskaran shortly after I got back: $T_c$ is limited by the kinetic energy which is reduced by the projection process,[iv] and vanishes when the doping *x* goes to zero.

I put together a few transparencies that same night, and in the morning asked the organizers if I could have a few minutes of session time. I gave the talk during the final session of the meeting.

Winging my way back across the Pacific, I was in a state of euphoria. I realized that I had merely connected dots, so to speak: each component of my idea had someone else's name on it. But really, the idea of an RVB, at least in this sense, that is surely completely my own, Pauling or no Pauling. No one else will have thought of it, and surely it's the only way to make this kind of superconductor. I will be famous and very much sought after, to quote Kipling's story about Yellow Dog Dingo. Be careful what you wish for--at this late date I certainly feel more like being chased by Yellow Dog Dingo than "famous and sought after".

# 3: A CORNUCOPIA OF IDEAS; AND THE FIRST, BUT UNNOTICED, PROGRESS

At this point the theory of the high-$T_c$ superconductors ceases to be a chapter in my autobiography and becomes much more suitably described in terms of a technical review article. Not your conventional one—I shall not cover all the bases and will not even mention most of the populous and popular red herrings that kept appearing, such as "anyon superconductivity", SO(5) theory, electron nematics, Eliashberg analyses, alternative orderings, spin fluctuations, etc, and I will only barely mention my own red herring, the interlayer tunneling theory, which was, deservedly, never either populous nor popular.

In the first months of effort by our group in Princeton and by others who had adopted the RVB idea, we learned some things of eventual importance. One was negative: the RVB is not, at zero doping, stable against antiferromagnetism, though the latter is rather weak and has a low $T_c$, and the best RVB is almost as low in energy. A second is that the BCS representation of the pure RVB is overcomplete: Baskaran pointed out that there is an enormous local SU(2) gauge group connecting different representations of the same wave function, which follows from the fact that the two different Fermion operators $c^\dagger_{i\uparrow}$ and $c_{i\downarrow}$ both create an up spin at site $i$ in the projected state. Finally, several groups concluded that the lowest energy RVB is (in one representation) the "$s+id$" state with two orthogonal gaps of shapes $\cos k_x \pm \cos k_y$, which state, however, is, in terms of electron coordinates, purely real and T-invariant (Shastry) (later we were to realize that nonetheless it has a $Z_2$ topological order). This state has an excitation spectrum which can be described by what I called "spinons", projected electrons, which have four point nodes in energy at the points ($\pm\pi/2, \pm\pi/2$) and extending upwards in Dirac cones. It turns out that the identical state in a totally different representation could be created by Gutzwiller projecting the single-particle state with a flux of $\pi$ through each plaquette (or, in fact, several other equally implausible ways.)

It was natural, though regrettably misleading, for Kivelson, Rokhsar and Sethna[v] to hypothecate the "holon", a charged object without spin which, combined with a spinon, made up an electron hole. Fractionalization was all the rage at that moment in time and we succumbed to that fashion, condemning me, at least, to a decade or more of a futile search for the holon. (I find a paper in *Science* as late as 2000 where I still claimed the electron fractionalizes in the cuprates.)

The next stage in the theory of the cuprate problem was carried out by two groups independently[vi],[vii]. Each arrived at the correct solution to the next crucial step, namely to what I called, much later, "spin-charge locking",[viii] and shall describe using the scheme of Ref. 6.

One way of expressing the SU(2) gauge group of the BCS-described RVB states is by using the Nambu-Anderson pseudospin algebra, i.e. to write the exchange interactions as

$$J_{12}(\mathbf{S}_1 \cdot \mathbf{S}_2 - 1/4) = J_{12}(\vec{\tau}_{12} \cdot \vec{\tau}_{21}),$$

where $\vec{\tau}_{12} = (\tau^1_{12}, \tau^2_{12}, \tau^3_{12})$ with

$$\tau^1_{12} = (c^+_{1\uparrow}c^+_{2\downarrow} + c_{1\downarrow}c_{2\uparrow})/2,$$
$$\tau^2_{12} = i(c^+_{1\uparrow}c^+_{2\downarrow} - c_{1\downarrow}c_{2\uparrow})/2,$$
$$\tau^3_{12} = (c^+_{1\uparrow}c_{2\uparrow} + c_{1\downarrow}c^+_{2\downarrow})/2. \qquad [4]$$

Both of these forms demonstrate that if there is no "*t*" term in the Hamiltonian, the energy depends only on the extent to which the bonds are singlets, since [4] gives zero for any triplet component. The energy for each bond is isotropic in the pseudospin space for that bond, which is the space which expresses the SU(2) gauge symmetry of the RVB states: clearly the τ-vectors can be rotated in this space in any desired way without affecting the exchange energy.

A simple physical interpretation of the "*s+id*" state is that it is the one in which all of the bonds parallel to the *x* axis—say—are singlet bonded with the maximum possible value of $\langle \tau^1 \rangle$ while the bonds parallel to the *y* axis have the maximum possible value of $\langle \tau^2 \rangle$, phased in such a way that the two directions do not interfere destructively. One may describe this state as a dyad of vectors in pseudospin space, which because of the symmetry is free to rotate keeping the two vectors perpendicular to each other (see Figs. 1—3).

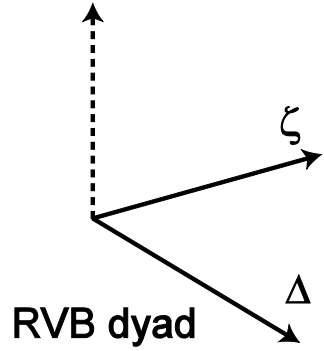

Figure 1  The true insulating spin liquid  RVB may be described in BCS representation by a dyad of vectors which are free to rotate in pseudospin ($\tau$-vector) space.

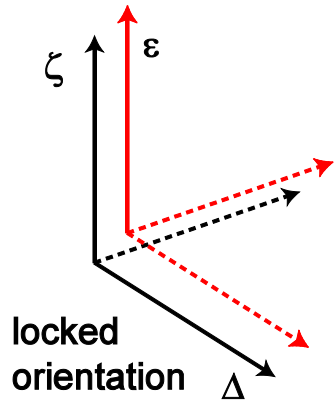

Figure 2   Minimizing kinetic energy, one arm of the dyad is locked to $\tau^3$ which represents charge.

The kinetic energy term can be reinserted by doping.  This term is definitely not isotropic in pseudospin space; the terms are unequivocally proportional to $\tau^3$:

$$T = P \sum_k \varepsilon_k (\tau_3)_k P \qquad [5]$$

and, in mean-field approximation, $P$ may simply be replaced by a renormalization factor $g_t$.  Both Refs. 6 and 7 made the obvious choice which pins down the rotational freedom of the pseudospin vectors.  The "*s*" gap function is identical in shape to the largest term in the kinetic energy, which comes from nearest neighbor hopping. Therefore, if one rotates the "gap" dyad so that (cos$k_x$ + cos$k_y$) multiplies $\tau^3$ (and assumes that the rotation is uniform for all bonds) that component of the gap and the kinetic energy will act together in their effect on the momentum distribution. In effect, this term in the energy "gap" due to pairing will give an extra component to the Fermi velocity which does not vanish as the doping *x* approaches zero. This term has been repeatedly confirmed experimentally[ix], and is often remarked upon by ARPES experimenters, who, however, do not refer back to its early prediction in Refs. 6 and 7.

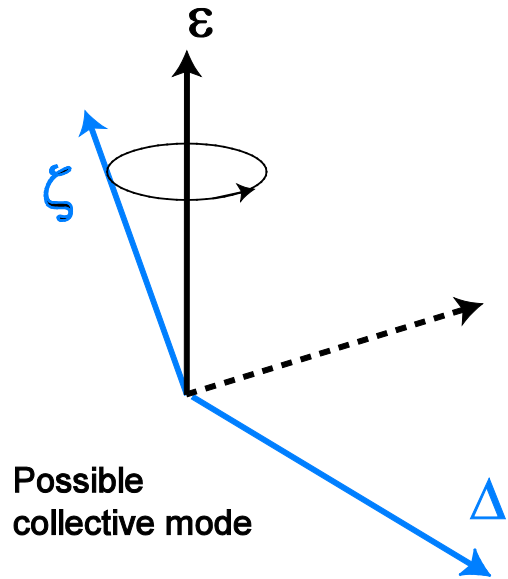

Figure 3  Illustrating that one expects at least one collective mode, which may be Raman-active, from oscillations of the locking mechanism.

The other component of the *s*+i*d* gap remains a true anomalous (superconducting) self-energy. Since the "spin-charge locking" mechanism fixes only the $\tau^3$ component of the gauge, the U(1) symmetry of rotations around $\tau^3$ is unbroken and makes the *d*-like component a true superconducting gap.

The most striking feature of this superconducting gap is that it is necessarily *real*. That means that the gap has zeroes at cos$k_x$ = cos$k_y$, where it changes sign, that are not filled in by a second imaginary component as is to be expected with angle-dependent gaps.[x] These zeroes are a very prominent feature in the cuprates, experimentally, since they contribute to thermal conductivity at low temperatures[xi], and particularly give a strong thermal Hall effect. I do not know an otherwise plausible mechanism for explaining these ubiquitous zeroes.

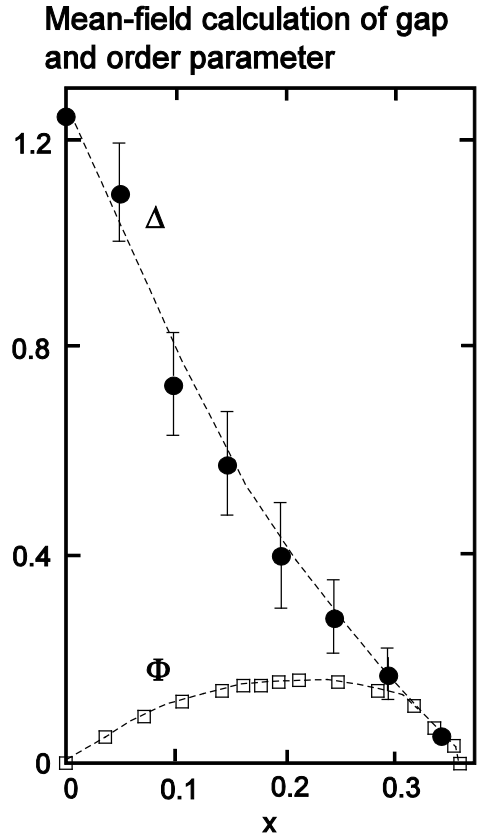

Figure 4  Calculation, from Ref 13, of energy gap and of true order parameter using the "plain vanilla" gap equations (Mohit Renderia, *private communication*).

In the original references 6 and 7, as well as in the pedagogical Ref. 8, it is assumed that the renormalization $g_t$ is the same for all **k** and throughout the sample. This may not be true for underdoped cuprates, as we shall see, but I know of no instance where the locking mechanism fails in an essential way. What does sometimes fail is our implicit assumption of homogeneity. Antiferromagnetism commonly occurs at the lowest doping and in the presence of magnetic fields; and charge density waves (stripes and checkerboards) are often seen at intermediate dopings. Both can be thought of as infrared instabilities of the basic RVB structure which establishes itself at relatively high energies.[xii]

Holding to the assumption of homogeneity, Kotliar and Liu, using a mean-field assumption in a "slave boson" theory, and Zhang *et al.* doing a quantum Monte Carlo, calculated nearly identical basic results for what they called the "renormalized mean-field theory"—results (I show in Fig. 4 a later version due to Paramekanti *et al.*[xiii]) in which the *d*-wave gap parameter rises to very high values (of order *J*) at low doping, while the superconducting order parameter, which is an estimate of the superfluid density, exhibits a "dome"-like behavior as *g* decreases to zero. All of these calculations preceded reliable experiments by at least half a decade, and the fact that these predictions were essentially borne out represents a triumph of what we came to call the "plain vanilla" theory—the simple Hubbard model transformed into the *t-J* model but otherwise unembellished by esoteric complications.

To return briefly to the autobiographical-historical mode: I, after the initial burst of creative chaos, chose in 1988-9 to take a disastrously wrong course, abandoning my original insight. I rejected the thought that the superconductivity could be "*d*-wave", reasoning that it appeared, qualitatively, to be relatively immune to disorder scattering, and searched for a mechanism which was capable of supporting *s*-wave superconductivity. Unfortunately, I seemed to find it in the "interlayer coupling theory" which placed great emphasis on the breakdown of Fermi liquid theory which was evident in transport experiments on the best-characterized cuprates. If, indeed, one ascribes this breakdown to the fractionalization of the electron into holon + spinon, coherent hopping of electrons between $CuO_2$ layers could become blocked and this could be meliorated by superconductivity, thus finding the energy for the phase transition in the interlayer hopping. I was brought to my senses in the years 1997-8, after one book, two courses of lectures, and innumerable meeting talks, by two impenetrable experimental facts. The first was the inexorable growth of evidence that indeed the order parameter was *d*-wave with nodes, and, especially, that it fit the cos-cos prescription for its **k**-dependence, as Mike Norman and J-C Campuzano determinedly demonstrated, especially via ARPES.[xiv] This means that it is *local* and almost certainly due to nearest-neighbor exchange. Notable steps in the demonstration of these facts were the early identification of *T*-linear dependence of the superfluid density by the UBC group of Hardy, Bonn and Liang[xv], ARPES gap measurements by the Stanford[xvi] and Argonne (ref 10 for example) groups, and finally interference measurements of the gap phase[xvii] which disposed of all other conjectures.

A second solid fact was the demonstration by Moler and Kirtley[xviii] that some genuinely high $T_c$'s, such as one-layer Tl cuprate, could be shown by direct penetration depth measurements in the interlayer direction to have extremely low interlayer coupling, far too low to provide enough energy for a 95 degree $T_c$. Superconductivity does occur in the isolated plane.

So I was dragged by the facts into the real world. In 1999 Mohit Randeria brought me into touch with Ref. 11, confirming and extending the work of Ref. 6 and 7 which I have discussed above. So many competing theories had, by that time, cluttered up the literature that it seemed necessary to write a review-type paper summarizing the ideas that many people around the world agreed on by now. In particular, the general acceptance of the *d*-wave order parameter had come at an unfortunate time, when among the most prominent ideas being touted was "antiferromagnetic spin fluctuations"[xix], based on a hazily-conceived analogy to the phonon mechanism of BCS, and to an extent to the ferromagnetic spin fluctuations which play a role—but not a crucial one—in He-3. Since such a theory necessarily had to assume *a priori* a nearest-neighbor antiferromagnetic exchange , any calculations would most probably lead to a *d*-wave order parameter and they did. Thus *d*-wave came to be seen as a vindication of the spin fluctuation "theory" rather than of the predictions of the RVB theory of Refs. 6 and 7, which determined the precise form of the gap—there had been even earlier suggestions that *d*-wave might appear, to be sure, notably a paper by Loh, Scalapino and Hirsch (Phys Rev B 34, 8190 (1986)), which seems to have anticipated Ref 19.

4: PLAIN VANILLA AND ANDERSON-ONG: THE MEAN FIELD THEORY FOR NEAR-OPTIMAL DOPING, AND THE ROLE OF ASYMMETRY

My idea was to put together a group of enough authors who agreed on the most basic elements of the physics of the cuprates that it would have some claim to represent a consensus view of the underlying problem. This is the genesis of the "plain vanilla" paper co-authored by myself, P A Lee, T M Rice, M Randeria, N Trivedi, and F C Zhang[xx]. These elements are [1] the superexchange vertex, resulting from [2] the projective canonical transformation, which [3] in turn renormalizes the kinetic energy and hence $T_c$ and the superfluid density. [4] The gauge ambiguity is resolved by the spin-charge locking mechanism. These ideas lead to the "plain vanilla" gap equations

$$\Delta(k) = g_J J \sum_{k'} \gamma_{k-k'} \frac{\Delta(k')}{2 E_{k'}}, \qquad [6]$$

with

$$E_k^2 = \xi_k^2 + \Delta^2(k), \quad \xi_k = g\varepsilon_k + \varsigma_k = g\varepsilon_k + g_J J \sum_{k'} \gamma_{k-k'} \frac{\xi_{k'}}{2 E_{k'}},$$

which explain some features of the data in detail, and others in a more or less broad-brush sense. Here $g$ and $g_J$ are renormalization factors for the kinetic energy $\varepsilon_k$ and the exchange energy $J\gamma_{k-k'}$, respectively, and $\zeta$ and $\Delta$ are the $s$-like and $d$-like gap parameters. The value of the gap parameter in Fig 1 approximately follows the vaguely defined pseudogap temperature $T^*$ at which a gap begins to open in the single-particle spectrum, while qualitatively, if not in detail, the superfluid stiffness controls $T_c$ which has its optimum value at a BCS-BEC crossover. For underdoping one expects a region of fluctuating "preformed pairs" between $T_c$ and $T^*$, in other words the "pseudogap" is qualitatively predicted (see Fig. 5).

The original intent for the "plain vanilla" paper was to make no claims to which any sensible person could disagree, in the opinion of the authors as a whole. We wanted the paper to represent, not a settled solution of the cuprate problem, but a progress report on where we had gotten so far, the things that could be agreed as a basis for further work. When we submitted it to the Reviews of Modern Physics, we were shocked to discover that although the editor was eager to publish it, he could find only a tiny minority of referees in the field who would agree. We had thought that the sociological situation was unfortunate, but we did not expect that that the field was so poisoned by self-interest and factionalism that no one outside our group could be found who wanted to see such a progress report published. In the end, we had to settle for a less prominent review journal with an editor willing to take a chance on us, and with a submission date delayed by a year.

With "Plain Vanilla" behind us, and in spite of the sociological trainwreck that was the search for "consensus", it seemed possible now to make progress in ferreting out the real physics without worrying oneself about competing views of the fundamentals.

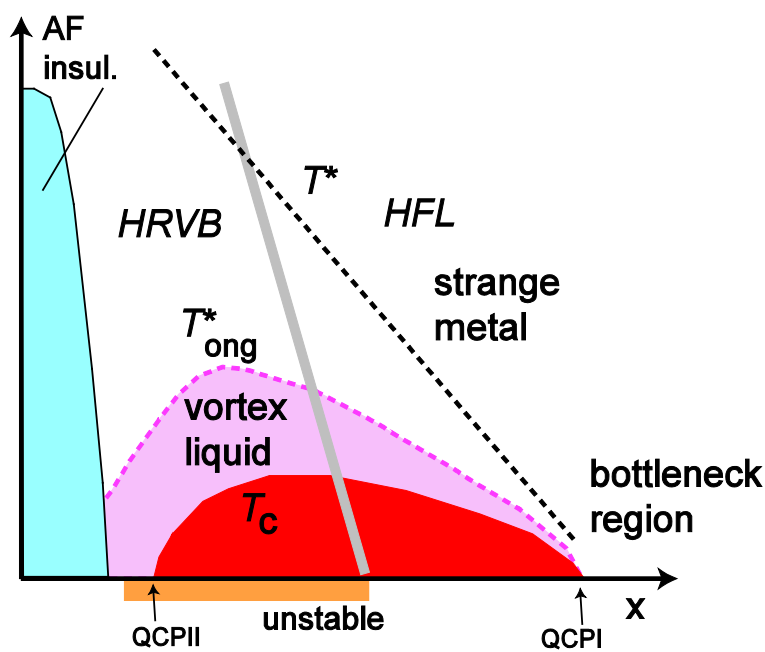

Figure 5  Schematic phase diagram of cuprate superconductivity. It is surprisingly possible to track features among the different materials, and in this sense it seems quite general.

Autobiographically, this was a period when two associations played an important role. First and most important was the little group we had here at Princeton, with Phuan Ong and his students, particularly Yayu Wang and Lu Li, (and later with Ali Yazdani and students), with V Muthukumar as an interlocutor and contact with Claudius Gros , his student Bernard Edegger, and others; and with my invaluable student Phil Casey.  Second was my regular attendance at the time at the Canadian institution CIAR, (now CIFAR) which brought me in touch among others with Seamus Davis and with the UBC group.  From discussions at CIAR grew my interest in the calculation of asymmetric tunneling spectra, from the former the characterization of at least a major part of the pseudogap phase as a vortex liquid and calculation of the vortex Nernst effect. Finally, with the discovery of how to underpin and to deal explicitly with Gutzwiller projection, came the full theory of strange metal transport processes (The hidden Fermi Liquid theory); and, following Maurice Rice and collaborators, I now work on the "hidden RVB", which bids to explain the weird behavior in the underdoped regime.

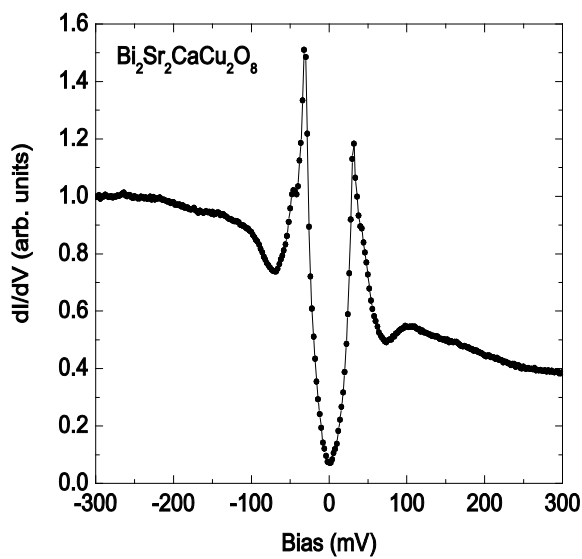

Figure 6   A typical STM tunnel spectrum for an optimally doped region of BISSCO (S H Pan, *private communication*).

Starting with Oystein Fischer's pioneering work with the STM in '95, [xxi] tunneling spectroscopy on some of the cuprates had been refined to the point where it was a trustworthy measure of the actual point density of states.  From the start in Ref. 21, and particularly clearly in  the work of the Berkeley group of Pan, Davis et al,[xxii] it was evident that the most consistent feature of these spectra was a marked asymmetry between the hole and electron sides, once one is at energies above the gap structure if that is present (see Fig. 6).  The asymmetry, as you would naively expect, favors hole tunnelling into the cuprate vs electrons, since holes can occupy many more sites.  But in the standard theory of tunnelling into metals, and in all previous experiments on ordinary metals, there is no such asymmetry because of a theorem due to Harrison.[xxiii]

Since this "theorem" has, like everything else of importance in the cuprates, become controversial, I reproduce the correct argument here. Bands in the cuprates, and particularly the $d_{x2-y2}$ one responsible for superconductivity, are well-represented by Wannier functions (else there could be no Mott insulator phase). Point-contact tunneling will take place primarily via the Wannier function directly under the tip. This Wannier function is precisely an equally-weighted sum of the Bloch functions of the entire band. The current in each of these Bloch waves is proportional to its velocity $\nabla_{\mathbf{k}}\varepsilon$, while the density of states is the inverse of this gradient; thus the band structure cancels out identically. Unfortunately, almost the entire theoretical literature on tunneling in the cuprates ascribes, implausibly, the asymmetric structure to the band structure, particularly to the evident van Hove singularity.

We start from the *t-J* renormalized Hamiltonian [2], the low-energy eigenstates of which must themselves come from the projected subspace: $\Psi = P\Phi(electron\ coordinates)$ where $\Phi$ is a function in the full Hilbert space. (We can accommodate a BCS function $\Phi$ if we use the fugacity factor—see below.) We take a different point of view towards the wave-functions [2] than has usually been considered: namely, we treat the problem as that of variationally determining the function $\Phi$ rather than the projected function $\Psi$: we seek the best energy obtainable by applying the projected Hamiltonian on to an *unprojected* variational function. We can avoid many difficulties by using the direct, equation-of-motion version of Hartree-Fock theory as described in my early book[xxiv]. We thus choose to approximate $\Phi$ by a generalized Hartree-Fock product function, that is the product of quasiparticle operators, one for each **k** and s:

$$\Phi = \prod_k (u_k c_{-k,-} + v_k c^*_{k,+})(u_k c_{k,+} + v_k c^*_{-k,-}) | vac >$$

$$= \prod_k \gamma_k \gamma_{-k} | vac > \qquad\qquad\qquad [7]$$

$$= \prod_k (u_k + v_k c^*_{k,+} c^*_{-k,-}) | vac > = \Phi_{BCS}$$

We define the ground state energy by

$$H\Phi_0 = PH_{t-J}P\Phi_0 = PE_0\Phi_0. \qquad [8]$$

Then we are going to determine the *u*'s and *v*'s by the Hartree-Fock procedure of demanding that all single-particle excitations have positive energies. That is we define the γ's by a set of Hartree-Fock-BCS equations:

$$\langle [H, \gamma^*_k] \rangle \Phi_0 = E_k \gamma^*_k \Phi_0 \qquad [9]$$

which are the plain vanilla gap equations we have already quoted.

In order to solve the tunnelling problem, it was essential to create a formalism for calculating Green's functions—and hence all physical responses—with Gutzwiller projected wave-functions; and in particular for tunnelling one must know Green's functions using states where an electron has been injected or removed. In order to do this correctly I borrowed an innovation of Bob Laughlin's[xxv], the fugacity factor. A BCS wave function is a wave-packet in total number *N* space, but that causes no difficulties because the packet is quite narrow and has its maximum at the physically correct number *N* for the chemical potential $\mu = E_F$. But Gutzwiller projection eliminates particles preferentially and distorts the wave packet, which maximizes at a very wrong *N*-value. This has no effect on the ground state or the gap equation [1], since the Hamiltonian conserves *N*, but is all-important to the single particle Green's functions. What is done is to multiply equation [3] by a "fugacity factor" which adjusts the probabilities of occupancies 0 and 1 to be *x* and (1-*x*)/2, rather than those of the original function which are in the ratio (1-*x*)/(1+*x*). Thus the new wave function to be projected is

$$|\Phi'\rangle = g^{-(n_\uparrow + n_\downarrow)/2} |\Phi\rangle. \qquad [10]$$

Here we recognize that the fugacity factor *g* is just the same as the renormalization factor $2x/(1+x)$ for *t*.

Equation [10] may be rewritten in a form which makes its effect much clearer. In each factor $(u_k + v_k c^*_{k\uparrow} c^*_{-k\downarrow})$ the *u* term is the amplitude for a pair of holes, the *v* term for a pair of electrons; so we can count up the number of hole pairs by just counting the number of *u* factors. Hence Φ' may be written

$$|\Phi'\rangle = \prod_k (g u_k + v_k c^*_{k\uparrow} c^*_{-k\downarrow})|0\rangle \quad [11]$$

Normalizing, we may write it in the form

$$|\Phi'\rangle = \prod_k (\tilde{u}_k + \tilde{v}_k c^*_k c^*_{-k})|0\rangle, \quad [12]$$

where $\tilde{u}^2 + \tilde{v}^2 = 1$, and

$$\tilde{u} = gu/\sqrt{u^2 g^2 + v^2}, \quad \tilde{v} = v/\sqrt{u^2 g^2 + v^2}. \quad [13]$$

We remind the reader that

$$u^2 = (1 - \varepsilon/E)/2, \quad v^2 = (1 + \varepsilon/E)/2.$$

The distinction between the *u,v* and $\tilde{u}, \tilde{v}$ is essential. *u* and *v* describe the effective unprojected states, and because they are used to compare states with the same number of electrons the fugacity factor is irrelevant, and *u=v* at the Fermi surface. $\tilde{u}$ and $\tilde{v}$ are parameters that describe the wave function in terms of real physical quasiparticles. $\tilde{u}$ and $\tilde{v}$ are parameters that formally describe the wave function. It is obvious from Eq. [11] that *g* renormalizes the order parameter and the superfluid density, since the contribution to each from any **k** is $2\tilde{u}\tilde{v}$. In the case of the superfluid density this is in accord with the standard sum rule argument which equates kinetic energy and superfluid density. In any singlet superconductor the pairs are partly pairs of spinless holes and partly singlet pairs of spins; what the *g* factors and the modified "cedilla'd" coefficients do is to allow us to change the relative admixture of the two. (The use of these coefficients and *g* is quite different in principle from the "Bogoliubov angles" of Balatsky[xxvi].)

Now to calculate the tunneling conductivity. We want to do this as simply as possible, so we set the factor $e^{iS} \approx 1$. The quasiparticle wave function may be created from the unprojected product function $\Phi'$ by either of the operators $\hat{P}c^*_{k\uparrow}$ or $\hat{P}c_{-k\downarrow}$. The former creates it with relative amplitude $\tilde{u}$, the latter with $\tilde{v}$. But what tunnels in from the metal is not a projected quasiparticle but a real one. The particle entering or leaving the layer comes into the Wannier function of the site immediately under the point; so the operators we must consider are $c^*_{i\sigma}\hat{P}$ and $c_{i\sigma}\hat{P}$.

Consider $c^*_{i\sigma}\hat{P}$. This may be divided into two parts, belonging to the forbidden and allowed subspaces.

$$c^*_{i\sigma}\hat{P} = (1-\hat{P})c^*_{i\sigma}\hat{P} + \hat{P}c^*_{i\sigma}\hat{P}. \qquad [14]$$

The first term can only create electrons in the forbidden subspace, with energy larger than $U$, and must be dropped. The second is finite only $x$ of the time, because the $c^*$ has to have found the site unoccupied. This factor $x$ is, of course, the primary source of the asymmetry, the reason why electrons find it more difficult to tunnel in than holes; but it is not the whole story.

Now consider the hole. $c\hat{P}$ automatically gives an allowed state. But $c_i\hat{P}|\Phi\rangle$ is not the same state as the corresponding combination of quasiparticle functions, $\hat{P}c_i|\Phi\rangle$, because the latter has components where the starting function $|\Phi\rangle$ contains $n_i=2$ components, while these don't appear in the former. Correspondingly, $c$ and $\hat{P}$ don't commute. We get

$$c_{i\sigma}\hat{P} = c_{i\sigma}(1-n_{i\uparrow}n_{i\downarrow}) = \hat{P}(1-n_{i-\sigma})c_{i\sigma}. \qquad [15]$$

We write this, dropping the site index,

$$c_\sigma \hat{P} = (1-\langle n_{-\sigma}\rangle)\hat{P}c_\sigma - \hat{P}(n_{-\sigma} - \langle n_{-\sigma}\rangle)c_\sigma = \frac{1+x}{2}\hat{P}c_\sigma + (inc). \quad [16]$$

The incoherent term I have written as (inc) involves the creation of three quasiparticles, and therefore will appear at higher energies. Estimates indicate that it will be relatively small of order $\Delta/t$, because the fluctuations of n extend over the whole bandwidth $\sim 4t$. But there is one interesting feature: because it contains terms like $\hat{P}[c*_{k-\sigma} c_{k'\sigma}]c_{k''-\sigma}$, which can excite the magnetic resonance + a quasiparticle, there is a bump in the amplitude near $\Delta$ + the energy of the magnetic resonance. This bump is not analogous to the "phonon bumps" of Rowell-Mcmillan.

Aside from this small term, Eq. [16] shows that in this approximation $c$, like $c*$, predominantly generates single quasiparticles in the superconductor; but the amplitude for doing so is $(1+x)/2x = 1/g$ times as large. In this sense most of the amplitude is coherent, a result quite different from previous authors[xxvii].

This ratio of amplitudes does not settle the question, however. A quasiparticle is a mixture of electron and hole amplitudes, and precisely at the gap it is an equal mixture, so that the conductivities for electrons and holes are identical. An electron tunneling in, for instance, can either result in an electron or in a ground state pair and a hole.

The calculation closely follows that of Tinkham[xxviii]. We use a tunneling Hamiltonian of which he writes the relevant terms as

$$\sum_{k,q} T_{k,q}\{a_{q\uparrow} u_k \gamma*_{k0} + v_k \gamma*_{k1} a*_{-q\downarrow}\}. \quad [17]$$

Here $T_{k,q}$ is the matrix element, which varies slowly with energy, $a_{q\uparrow}$ is the electron field in the metal point, and the $\gamma$'s are the quasiparticle operators in the cuprate layer. But in defining and normalizing the quasiparticle operators there is a subtle problem which was glossed over—though solved correctly—in previous versions of this work.
In writing the wave function [8] using the $\tilde{u}, \tilde{v}$ coefficients, with $\tilde{u}^2 + \tilde{v}^2 = 1$, and using standard BCS formalism hence using the quasiparticle operators $\tilde{\gamma}$ as defined in Eq. [3] with these coefficients, we are making a subtle error. These quasiparticle operators obey standard commutation relations

$$[\tilde{\gamma}^*{}_k, \tilde{\gamma}_{k'}]_+ = \delta_{kk'}$$

and behave therefore like quasiparticles in a conventional Fermi system, since—as we emphasized above—they are the *pre-projection* quasiparticles operating in the full Hilbert space before projection. Thus they are what, in discussing the Fermi liquid, we have called *pseudoparticles*—true excitations, but not continuously associated to the real electrons and holes of the system. In fact, if we were to directly use them in deriving the tunneling spectrum, we would find it symmetric.

The true quasiparticles have finite overlap with the pseudoparticles, in the superconductor, because their decay channels come with finite energy gaps. In the mean field approximation, we neglect completely the incoherent, inelastic effects of the three-Fermion terms, and in this approximation a pure hole at high energy encounters no QP-PP barrier. On the other hand a pure electron is relatively obstructed by the universal factor *g*. A quasiparticle, which is a mixture of the two in the ratio $gu \div v$, will act as a transmission channel which transmits with the amplitude

$$T = \sqrt{(g^2 u^2 + v^2)} \qquad [18]$$

and we must multiply by this overall factor. Finally, the tunneling density of states for electrons turns out after putting all of this together[xxix] to be

$$N_e(E, \Delta) = \frac{d\varepsilon}{dE} g \left( \frac{u^2}{\sqrt{u^2 + g^2 v^2}} + \frac{v^2}{\sqrt{v^2 + g^2 u^2}} \right). \qquad [19]$$

The *g* factor in this formula comes from the projection factor [18]—which is essentially the transmission factor of the channel between quasiparticles and pseudoparticles. As we see, for $v \cong 1$ at high voltage, the limiting value, 1, comes from the second term and the tunneling is suppressed by *g*. On the other hand, for holes the tunneling density is

$$N_h(E,\Delta) = \frac{d\varepsilon}{dE} g\left( \frac{v^2}{\sqrt{u^2 + g^2v^2}} + \frac{u^2}{\sqrt{v^2 + g^2u^2}} \right). \quad [19]$$

Here the *g* factor comes from the normalized fugacity factor, and at high voltage $u \cong 1$ and *g* cancels out, giving the ratio *g* between the two limits. It is striking that in many quite different spectra, where the details of the gap-region spectra have been complicated by inhomogeneities, structure, etc, the asymmetry at 100 meV or so seems to provide a good estimate of the doping level.[xxx] The remarkable, but subtle, symmetries of the two expressions are striking.

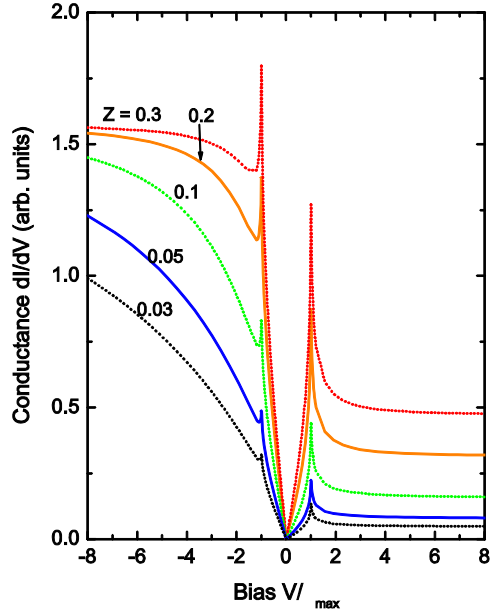

Figure 7  Tunnel spectra calculated as described in the text for different values of the asymmetry parameter $Z = 2x/(1+x)$ (the same as *g* in the text).  The very asymmetric limit, as $Z = g \rightarrow 0$ is noteworthy.

These formulas must be integrated over the *d*-wave distribution of gap values to give a prediction for comparison with observed spectra. This we have done only roughly, using $P(\Delta) = 1/\sqrt{1-\Delta^2}$, as though the Fermi surface were circular and not taking into account the actual band structure, which does somewhat affect the distribution of *Δ* values. In Fig. 7, we give the predicted spectra for a number of values of *g*, using this simplification.

It is obvious that observed spectra for optimally doped patches, as illustrated in Fig. 6, have perceptibly more prominent coherence peaks than the prediction, but otherwise the *general course of the experimental results is in fair agreement with this mean field theory for dopings not too far from optimal, and in excellent agreement as far as the unique features of the spectra are concerned: the "hump and dip" occurring primarily on the hole side, at energies related to the gap and the doping percentage; their absence on the electron side; and the symmetry of the singularity at the gap energy. We cannot emphasize too strongly the support that this agreement gives to the general theoretical picture behind it. The asymmetry is incompatible with any perturbation method starting from weak coupling (as for instance spin fluctuation theory, or any phonon-based theory,) and shows that the system **must** be described as a doped Mott insulator.*

5: THE OVERDOPED CASE: HIDDEN FERMI LIQUID AND THE BOTTLENECK

At this point I abandon chronological order and go next to the overdoped case. I imagined the energy gap vanishing, as it does as one increases the doping past the *T\** line on the generalized phase diagram (Fig. 5) at which pairs form. In that event, the mean field result of Anderson-Ong implies a jump discontinuity at the Fermi energy. This in turn implies an unacceptable logarithmic divergence of the self-energy at $\omega=0$[xxxi] and forces a resummation of the divergent diagrams. Fortunately this problem turns out to be identical with the old problem of the "*x*-ray edge singularity" which was solved in the early 1970's by Mahan,[xxxii] Nozieres and coauthors[xxxiii], and myself[xxxiv].

But perhaps more important was that this problem forced a conceptual rethinking of the methodology. I began to realise that the Kohn-Rice canonical transformation to a 100% Gutzwiller-projected Hamiltonian was the fundamental step in creating an effective low-energy theory for the Hubbard problem.

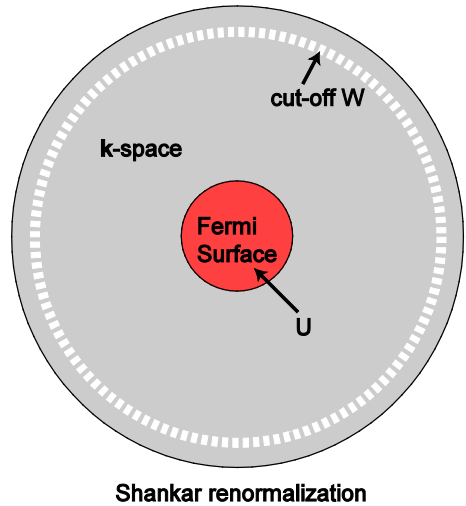

Figure 8  Starting point for "poor-man's" renormalization: kinetic energy > $U$, squeeze all interactions down.

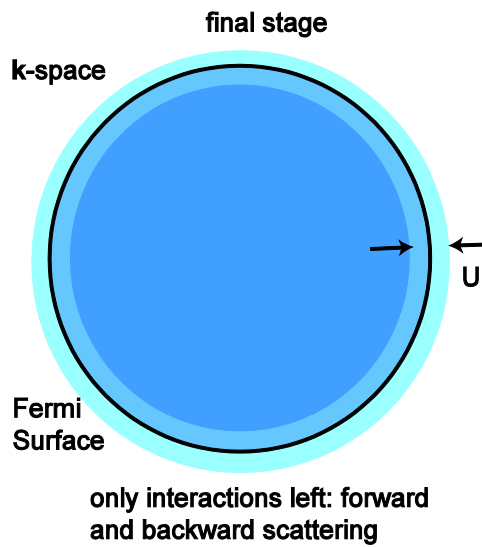

Figure 9  The final stage: all relevant states are in a shell around the Fermi surface.

A paper from the 2004 Sitges meeting [xxxv], (referring back to a review by Shankar explaining the "poor-man's renormalization group"[xxxvi]) makes it clear that any Fermion system can be reduced to a low-energy theory where all the relevant degrees of freedom are confined to a shell close to the Fermi surface **iff** the Fermi energy is the largest interaction in the Hamiltonian. But exactly this is not the case for the strong-coupling Hubbard model, where $U$ is large enough to cause the Mott insulator phenomenon by dominating the Fermi energy.

Thus it is necessary, in creating a low-energy effective theory, to *first* carry out the essential step of removing $U$ from the problem by "renormalizing it to $\infty$". (see Figs. 8, 9 and 10) The result is the *t-J* Hamiltonian [2] (or its equivalent in more complicated cases.) Thus in principle it is not optional to work with the projected *t-J* Hamiltonian.

So much is pretty much rigorous. But the next step is constrained by necessity: we don't know any other scheme than to use as a trial function a product function like [12] of single-Fermion operators. And the other, unique, procedural concept is that by doing so, we are free to use the **projective** *t-J* Hamiltonian to determine variationally the formally **unprojected** product function, because momentum eigenstates are automatically orthogonal.

But this *Ansatz* does not solve the whole problem. Unfortunately there are different solutions to this variational problem. The *t-J* Hamiltonian has two terms, and there are three possible situations, depending upon doping and temperature: [1] $t$ may dominate, as is likely at high doping; [2] both terms are relevant, but $J$ may be treated as a perturbing interaction; or [3] $J$ is dominant, and determines the shape of the spectrum, but $t$ is also relevant.

Case [2] is what I have already discussed above, and treated in the Anderson-Ong paper, Ref. [27]. Because it develops a true energy gap, the optimally doped superconductor can be treated to a fair level of approximation as a simple modified BCS theory ignoring the incoherent terms caused by projection.

Case [1], which applies everywhere above and to the right of the pairing line $T^*$, is more complicated because there is no energy gap. $J$ may be taken into account merely by renormalizing Fermi velocities and allowing for possible minor modifications of the Fermi surface shape, but the crucial *Ansatz* is that there **is** a Fermi surface, i.e. that the projected kinetic energy, acting on fermions in the unprojected space, exhibits a sharp locus in **k**-space of zero-energy single-Fermion excitations and a spectrum of excitations whose energies rise linearly with $|k–k_F|$. These excitations would decay by Umklapp processes at a rate proportional to $|k-k_F|^2$. Such an *Ansatz* is obviously self-consistent, by the same arguments based on momentum and energy conservation which underpin the conventional Fermi liquid, but we have no rigorous proof that it is true. The best evidence that it is is experimental: fitting laser-excited ARPES energy distribution curves to shapes derived from this assumption (see Fig. 11[xxxvii]).

These single-Fermion operators acting within the unprojected Hilbert space are the "Hidden Fermi Liquid." They represent true eigenexcitations of the system but they are not true quasiparticles—we designated them "pseudoparticles". We designate the "real" Fermions which represent physical creation and destruction operators acting in the projected subspace by "hat" operators which do not create or destroy any doubly-occupied sites. These are easily seen to be

$$\hat{c}_{i\sigma} = (1 - n_{i-\sigma})c_{i\sigma} \quad (\hat{c}^*_{i\sigma} = c^*_{i\sigma}(1 - n_{i-\sigma})) \qquad [20]$$

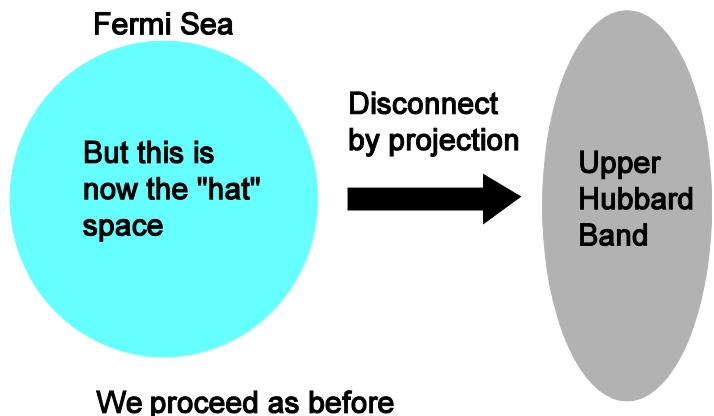

Figure 10 But if $U$ > bandwidth, it doesn't work: instead, $U$ must be removed by projection.

That is, the quasiparticles are three-pseudoparticle operators.

They are the physical particles which are exchanged by tunneling or acted on by physical fields. It is the Green's functions of the true quasiparticles which underlie all of the phenomena of the "Strange Metal" (sometimes referred to as the "Marginal Fermi Liquid") region of the cuprate phase diagram. As I explained in the (progressively more accurate) references[xxxviii,xxxix], the Green's function is to be calculated by factorizing the three-pseudoparticle operator into two dynamically independent pieces in two ways.

$$\hat{c}_{i\sigma} = c_{i\sigma}c_{i-\sigma}c^*_{i-\sigma} = c_{i-\sigma}S^-_i \text{ or } = c_{i\sigma}(1-n_{i-\sigma}) \qquad [21]$$

The quasiparticle near the Fermi surface can thus decay into pseudoparticles in two different ways, with or without flipping the spin of the residual pseudoparticle which carries most of its Fermi momentum. The bosonic operators $S^-$ and $(1-\rho)$ have an infrared singularity in their decay rates, as a consequence of the "infrared catastrophe"[xl] which plays such a role in the "x-ray edge" problem and the Kondo problem (see ref 32); as a result their Green's functions decay with a power law $t^p$ at $T=0$, which at finite temperatures becomes proportional to $(T/\sinh \pi T t)^p$.

This is responsible for the "linear $T$" relaxation rates which are such a striking feature of the strange metal. One other slightly unfamiliar idea completes our understanding of this region: the "bottleneck". A great deal of experimental data over the years has supported the idea that there are two relaxation rates and two relaxation processes active in this region, characterized by linear $T$ and quadratic $T$ relaxation rates. Most recently, the extensive and detailed data of Hussey et al[xli] bring this out explicitly. But the full array of data has not been rationalized with any of the conventional schemes which have been tried, specifically the "hot spot" idea that they belong to different sectors of the Fermi surface. When we realise that these are two qualitatively different relaxation processes acting in series to carry the momentum supplied by the field to the lattice, we understand the process in a new way. The quasiparticles are the entities accelerated by the electric field; but they are not even approximate eigenexcitations and they must decay into pseudoparticles before the umklapp transitions which relax the momentum to the lattice can operate.

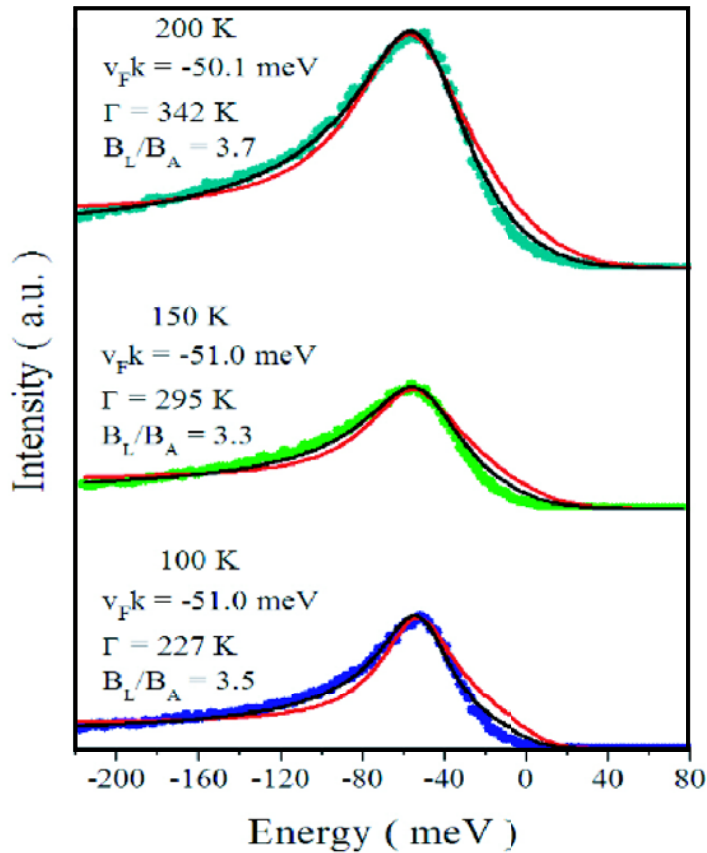

Figure 11 Fits to laser-excited ARPES energy distribution curves at $T>T_c$. The asymmetric Doniach-Sunjic curves (black) are better fits than Lorentzians (red) with NO added background and one fewer fitting parameter.

The linear $T$ quasiparticle decay is momentum-conserving and does not in itself cause resistivity; it actually *protects* the quasiparticles from scattering because the singularity reduces the density of final states for quasiparticles. Thus the process is two dissipation processes in series, which is equivalent to two conductivities in series, not two resistivities. A general formula for the resistivity, then, would be

$$\rho - \rho_{res} = \left[1/\rho_{qp} + 1/\rho_{umklapp}\right]^{-1} = const \frac{T^2}{T+W} \qquad [22]$$

It is remarkable how well this fits data from a wide variety of sources. I emphasize that I have seen no data indicating a transition to a true Fermi liquid. Using these two concepts, Casey[xlii] has been able to understand the full range of data on the bulk properties available in the strange metal region.

The bottleneck process can explain one striking but not often remarked fact about the resistivity: that the "linear $T$" resistivity often passes without a hitch right through the Mott-Joffe-Regel "maximum metallic resistivity" without saturating as conventional phonon resistivity in metals does.

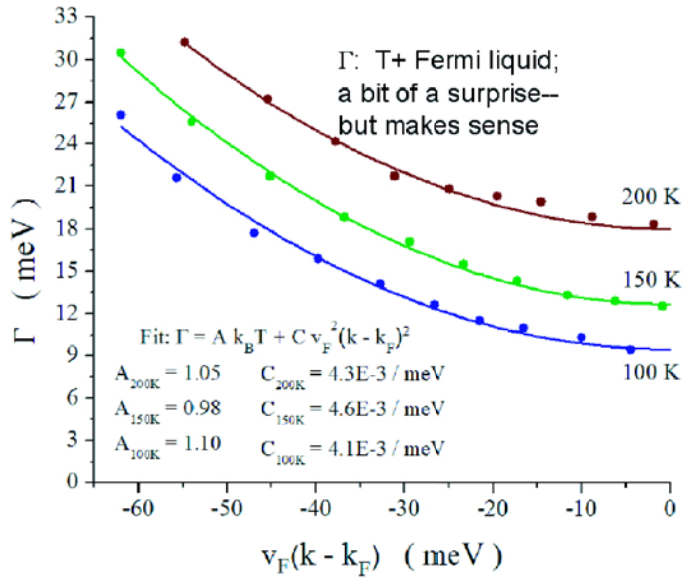

Figure 12  Parameters for fits which reveal the linear-$T$ and $(k-k_F)^2$ components of the HFL picture.

6: UNDERDOPING: THE HIDDEN RVB?

Now we must attack the truly complicated case [3]: the $J$ term dominates and kinetic energy is a weak perturbation. In this case we must start from a solution for the $J$ term alone as a basis, to establish an excitation spectrum analogous to the Fermi surface of case [1]. Of course, for the true insulator that is the simple antiferromagnet, but experiment tells us that a tiny bit of doping or a moderately low temperature destroys spin ordering, and what seems to take over is the "$s+id$" RVB structure.  In any case the two are actually very similar in their correlations and the higher $T$, higher energy behavior will be the same. But the initial expectations that the RVB pairing would go over smoothly into a superconducting state were naïve, the RVB is initially an insulator on two counts: first, that with weak kinetic energy, i.e. small effective *t*, the spin-charge locking mechanism doesn't work and the pairs are effectively neutral; second, there is a commensurability energy which holds the structure in register with the crystal lattice: that is, the nodes are pinned at the π/2,π/2 points.[xliii]

As we add in holes, we pin the chemical potential for holes near the node energy, and the holes are added as "conventional" pseudoparticles, not as holons, forming Fermi surface pockets near the π/2,π/2 points as proposed by Rice *et al* in Ref. 43; the RVB structure does not change and provides the major proportion of the pseudoparticle self-energy. (See Fig 13, borrowed from Ref. 43.) But the pairing self-energy from the insulating RVB structure does not have the U(1) rotational freedom in τ-space of a true anomalous superconducting self-energy; and Ref. 43 introduces the appropriate self-energy heuristically, without any accompanying superfluid responses.

One way to formalize what might be going on here is to take advantage of the freedom inherent in the type of wave function [11] or [12] where the pair amplitude is allowed to vary from pure spin singlet pairs to pure spinless hole pairs. We cannot find a reasonable argument which prevents us from allowing the parameter *g* to vary from place to place on the "Fermi surface". The constraint on a putative $g(\mathbf{k})$ is merely a global one, that it should lead to a correct number of electron holes. $g=0$ everywhere is the pure Mott insulator RVB, which makes no accommodation to the kinetic energy (and has no superfluid density $\rho_s$) and optimizes the response to *J*, while $g=1$ is the straightforward metallic superconductor, which has the maximum possible response to the kinetic energy consistent with the pairing gap. I hypothesize that near the nodes *g* takes on a value near to 1, since here the gap is small and it is important to optimize the kinetic energy. The hole excitations on the nodal lines will feel no or a small gap and will show the putative Fermi surface for the ordinary kinetic energy. Reference 43 shows that such excitations form hole pockets as shown in the Fig. 13. Near the antinodes the gap overwhelms the kinetic energy and $g \approx 0$. The pairing persists, but it is only pairs of pure spins with no hole pair component.

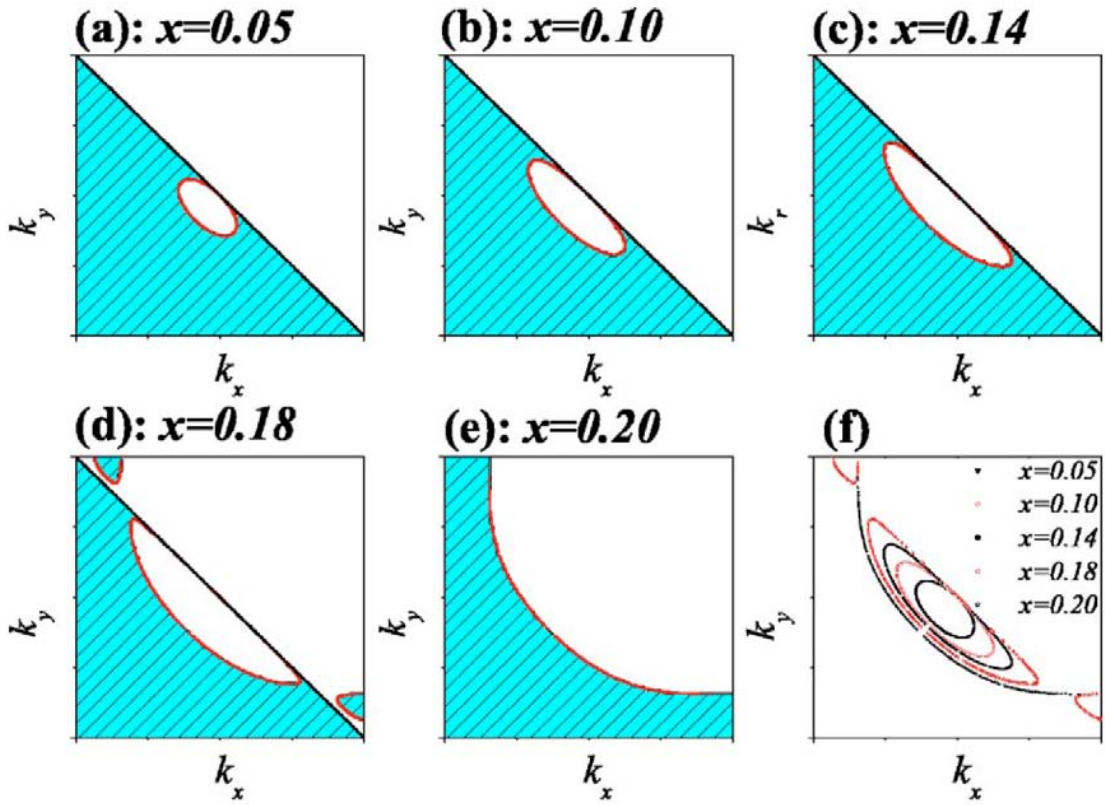

Figure 13  Hole pockets in the picture of Ref 43.

There seems no reason for *g* to be continuous, and indeed there is a prominent "kink" in the tunneling spectrum (according to A Yazdani) which may be the location of the changeover in the value of *g*. Possibly a branch of pure spinon excitations persists all the way to the nodes. But true superconductivity, when it develops, is confined at first to the pocket-arcs alone.

This idea is so far only a conjecture. I have not fleshed it out with a formal theory.    But it seems to correlate with all the known experimental facts, including the "Fermi arc" phenomenon[xliv] and the mysterious simultaneous existence of strong Dirac nodes and of hole pockets.[xlv] Any detailed discussion should await a full paper.  But clearly the correct approach to this region of the phase diagram is to think in terms of a "hidden RVB", as for the highly doped case it is the "hidden Fermi liquid".

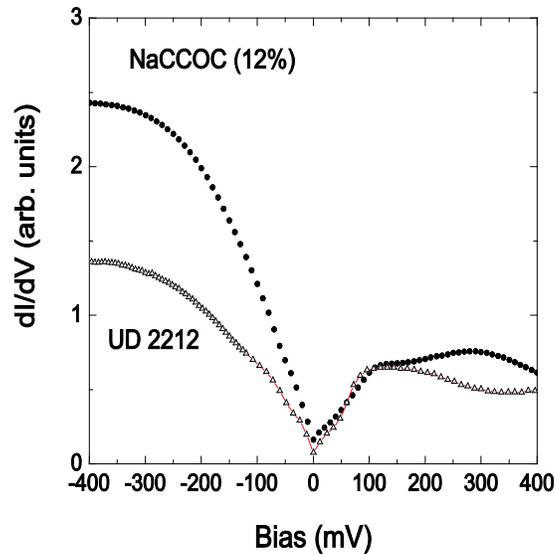

Figure 14 Extreme asymmetry and "kinks" in tunneling spectra of very underdoped cuprates showing dichotomous behavior for low and high energies. (J C Davis, *private communication*).

It is particularly hopeful that this structure can be seen in the tunneling spectra of severely underdoped cuprates—see fig 14, from Davis *et al.*— where the high-energy spectrum is completely one-sided, as expected from an insulating RVB, but at lower energies breaks sharply into a symmetrical *V*.

7:THE VORTEX LIQUID

In the course of his transport studies of the cuprate superconductors N P Ong and his students came upon an extensive region of the *x,T* phase diagram in which these studies gave unequivocal evidence of quantized vorticity above the superconducting $T_c$. The initial discovery was via the Nernst effect, the generation of an electric field perpendicular to crossed magnetic field and temperature gradient, interpreted as the field generated by thermally driven vortices[xlvi]; and it has since been confirmed by measurements of non-linear diamagnetism with an ingenious technique.[xlvii] This equilibrium effect can be shown to be equivalent to the Nernst measurement if and only if the former is due to vortex motion. The region in which this occurs is shown for two sets of compounds in Fig 15. The facts that these phenomena occur over a finite, in fact extensive, region of phase space, and that one manifestation is a non-analyticity of an equilibrium property, argued to us that this was not critical fluctuations (as proposed in[xlviii]) but a distinct new phase of matter.

This idea is supported by models for $T_c$ such as the generally popular 2D Kosterlitz-Thouless one, where $T_c$ occurs not by the BCS breakdown of pairing but by loss of phase rigidity due to proliferation of vortices, with no implication of vanishing of the superfluid density $\rho_s$ at the microscopic level. The same "XY" model principle was proposed for HeII by Feynman and others, and followed up by G Williams;[xlix] and it is generally agreed that the cuprate $T_c$'s tend to resemble 3DXY critical behavior.

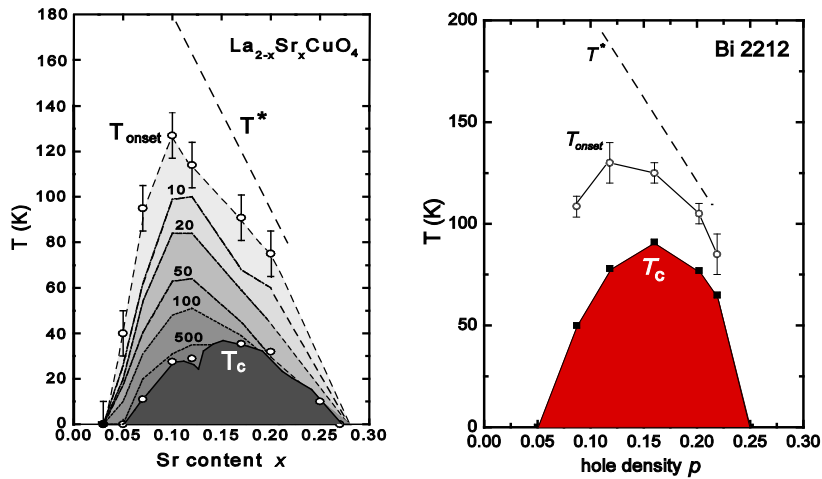

Figure 15 The "vortex fluid" region in which a vortex Nernst effect is measurable, in two compounds with a range of dopings (N P Ong, *private communication*).

The diamagnetic susceptibility, which is the response $\partial^2 F/(\partial B)^2$ to the vorticity in the electron gas induced by a magnetic field $B$, is relatively large and non-linear at field scales which are relatively low—in some cases it is even divergent at low $B$. On the other hand, the resistivity is finite and perfectly linear.

In the phase space region which Ong investigates, the electrons are thought to be fairly strongly paired, superconductivity having been destroyed only by phase fluctuations of the order parameter, as demonstrated in 1993 by Salomon[l]. Therefore it is reasonable to think of the currents as being predominantly carried by paired electrons, i e bosons. If there is a finite local pair amplitude above $T_c$, the pair wave function will have a time- and space-varying phase $\Phi$ and the pair current will be proportional to $\nabla\Phi$ and conserved. If so, $\Phi$ will be completely determined by a network of vortex lines — in 3D, mostly vortex loops. Thus it is appropriate to describe this phase as a vortex fluid.[li]

We will now try to make the existence of this state theoretically plausible. To do so I will revert to the 2D model of Ref. [6], although I believe that the results generalize simply to 3D. The current in a 2D system of vortices is simply the sum of those due to the individual vortex points: (We scale $\rho_s$ to 1 for convenience.)

$$J_i = \nabla\Phi_i = q_i \hat{\theta}_i / |r - r_i|$$
$$q_i = \pm 1 \qquad [23]$$

There must be a lower cutoff a around the vortex points if only because the velocity can't be infinite; this will be implicit in all further work. The energy is then the integral of the square of the sum of all the contributions in Eq. [23]:

$$U = \frac{1}{2} \int d^2 r (\sum_i J_i)^2 \qquad [24]$$

The integration in Eq. [24] may be carried out and the result is, introducing an upper cutoff radius $R$ for the sample as a whole which is more or less identical for all vortices:

$$U/2\pi = \sum_i q_i^2 \ln(R/a) + \sum_{i \neq j} q_i q_j \ln R/r_{ij}$$
$$= (\sum_i q_i)^2 \ln R/a - \sum_{i \neq j} q_i q_j \ln r_{ij}/a \qquad [25]$$

We note from the first line of Eq. [25] that each individual vortex has a self-energy which diverges logarithmically as $2\pi\ln(R/a)$; but that if the system of vortices is neutral with $\Sigma_i q_i=0$, the dependence on sample size cancels against terms from the sum of all the other vortices and the standard Kosterlitz-Thouless interaction energy results, with no dependence on the upper cutoff radius:

$$U_{K-T} = -2\pi \sum_{i \neq j} q_i q_j \ln(r_i - r_j)/a \quad + \sum_j E_c \quad [26]$$

The core energy $E_c$ for the local energy cost of a point zero of the boson field could formally be subsumed into $a$.

If, however, there is a mismatch in the + and - vortex numbers, there remains a divergent term proportional to the logarithm of the upper cutoff radius. This term is formally proportional to the square of the mismatch but there are also long-range uncompensated terms in the interaction in Eq. [26], and taking these into account it turns out that effectively we must add in a large self-energy $2\pi\ln(R_B/a)$ for each unpaired vortex, if their distribution is reasonably uniform. $R_B$ is approximately the distance between unpaired "field" vortices. **This term has been omitted in all previous treatments of the "normal" bose fluid, as well as in discussions of the superconducting vortex fluid.[lii]** It is crucial to the Abrikosov theory of the vortex lattice, but its implications for the normal state have not been explored.

A mismatch in vortex numbers means that the sample has net vorticity, i.e. is rotating as a whole (or, in the superconducting case, that it is experiencing an external $B$-field). As has been understood since the '50's[liii], the minimum energy configuration will be a uniform array of vortices, which is the closest mimic of rigid rotation. At length scales greater than the distance between unmatched vortices $R_B$ the physics is macroscopic and classical, and the quantization of vorticity is irrelevant. In this regime the divergent self-energy for $r>$the lattice constant of the array may be cancelled against whatever source of energy is causing the rotation or against the source energy of the $B$-field.

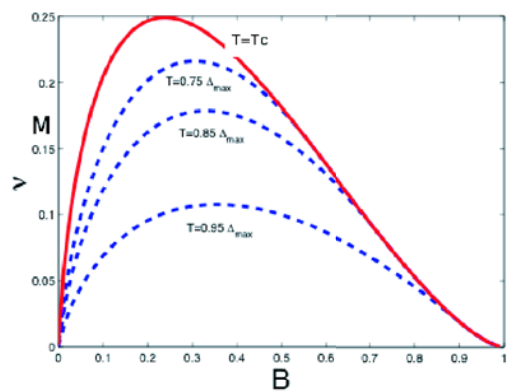

Figure 16  The predicted shape of the vortex energy ---i e Nernst effect or planar diamagnetism –vs field at $T_c$ and above.

But there still remains the energy caused by quantization of the vorticity, leading to a nonuniform local velocity. This energy is (if the density of extra vortices is $n_V$) proportional to

$$n_V \ln(R_c^2/a^2) = n_V \ln(1/n_V a^2) \quad [27]$$

$n_V$ is constrained by the need for canceling the divergent terms to be proportional to $B$ for the superconductor (or to $\omega$ for a superfluid.)

The crucial point is that the energy Eq. [26] *is not screened out by the thermally excited pairs above $T_c$.* This is counterintuitive relative to one's experience with the apparently similar system of electrically charged particles; but it is true. One important difference between the two cases is that the self-energy of charged particles is ignored as being compensated by their chemical potential, whereas here it is part of the dynamical problem.

The result may be understood by simply examining the energy expression Eq. [25]. This consists of the sum of all interaction energies, each term proportional to $\ln(r_{ij}/a)$; and the self-energy term, which is independent of all $r_{ij}$'s and depends on $\ln R$. Adding an unmatching vortex cannot change the interaction sum by this much: at a distance, the extra vortex encounters a neutral gas of vortices, and close by, it may have attracted a screening cloud— consisting of a single quantum of vorticity-- at a radius of order $1/(n_{pairs})^{1/2}$, but that doesn't give a large term in the energy. When a quantum of vorticity is added, its circulating current is uncorrelated with the others' at distances less than $R_c$. We may think of this temperature region as being dominated by entropy, leaving the vortices quite uncorrelated but uniformly distributed.

In 3 dimensions, the computation is more complex but if we are above the XY model $T_c$, entropy dominates and the extra vortex line is not effectively screened by vortex loops.

On the other hand, Eq. [26] is the energy, not the free energy. Below $T_c$, it is controlling and gives us, for instance, the Abrikosov theory of the vortex lattice in superconductors. Below $T_c$ the thermally excited vortices are bound in pairs and partially screen the interactions. In the Kosterlitz-Thouless theory, $T_c$ occurs where the extra logarithmic energy of free vortices is compensated in the free energy by $T$ times the logarithmic entropy which one gains by allowing the vortex to be anywhere in the sample. Above $T_c$, pairs of vortices proliferate in the critical region in such a way that their number is given by the activation expression which results from equating the logarithmic terms in energy and entropy:

$$n_{pr} = (1/a^2)\exp[-E_c/(T-T_c)] \qquad [28]$$

Interactions reduce the numbers of vortices but do not cause strong correlations among them.

The large extra entropy of the unmatched vortices does not cause them to proliferate because they are indistinguishable from the positive members of pairs whose number obeys Eq. [28]. Actually, the free energy of order $n\ln(n)$ cancels exactly, as we can deduce by using the identities $U=\mathbf{M}.\mathbf{B}/2$ and $M=\partial^2 F/\partial B^2$. This cancellation does not occur to higher order in $n_V$ leading to a free energy term of form $(n_V)^2\ln(1/n_V)$, which gives the logarithmic response function.

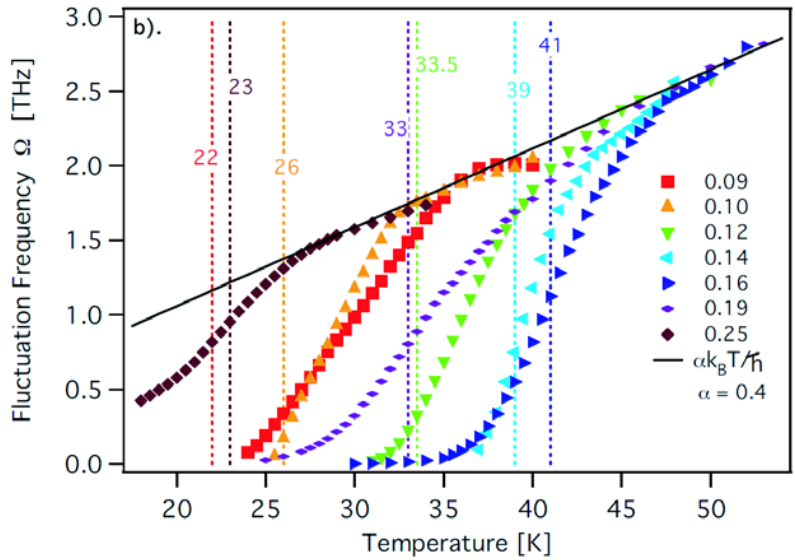

Figure 17 Relaxation rate as measured by terahertz spectroscopy vs $T$ for variously doped LASCO samples, showing critical and linear-$T$ regions. [Armitage, Ref. 56]

The Nernst effect gives us a uniquely direct way of measuring specifically the energy carried by vortices. This is because of the reciprocity between the Nernst effect—the voltage response to a heat current—and the Ettingshausen effect, the heat current response to a given voltage. An $E$ field implies that the net vorticity moves at the velocity

$$v = cE/B \qquad [29]$$

The heat transported is just the actual energy Eq. [27] of the vortices, so the Ettingshausen coefficient $\alpha_{xy}$ is proportional to $B\ln B$. The actual shape of $\alpha_{xy}$ vs $B$ is complicated by the strong dependence of $\rho_s$ on $B$, and has been discussed at length and quantitatively elsewhere.[liv]

An equally challenging experiment is the direct measurement of the vortex energy via the magnetization. The energy must come from the interaction of the current with the field, which is **J**·**A**/2 or, equivalently, **M**·**B**/2, so the diamagnetic moment is a direct measurement of the energy due to the added vortices. When, as is often the case, the two measurements (Nernst effect and diamagnetism) yield nearly identical results related by the factor $2/T$, that is strong evidence for our interpretation in terms of quantized vortices.

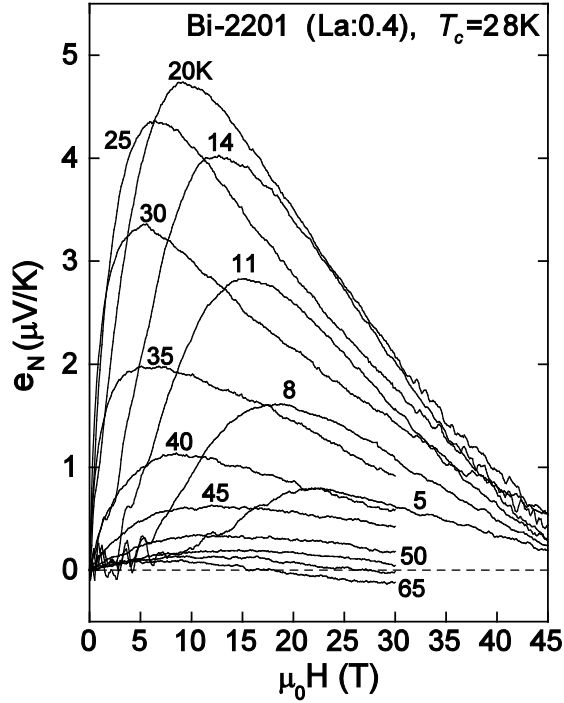

Figure 18  Nernst effect vs field for an underdoped 1-layer BISSCO (Ref. 46).

Both the Nernst effect and the magnetization can have other causes. There are several ways in which a Nernst effect can result from particle currents, although except in special circumstances these effects are small and linear.  It is also clear that magnetization can result from  spin susceptibility or Van Vleck paramagnetism, which are completely independent effects.  Thus identifying the vortex term by using the two measurements together may often allow a fairly unequivocal diagnosis of vortex motion.

 Most treatments of the Bose liquid above $T_c$ such as Ref. 49 have restricted themselves to the critical range near the λ-point or K-T transition. (Also see [lv]) But as we see, the anomalous response is not a critical phenomenon but an intrinsic property of the vortex liquid phase, and should persist as long as there is a finite core energy for vortices. In Ong's Nernst effect fluid there seems to be quite a range above the critical region which is characterized by a correlation time for vortex flow of around $\hbar/kT$, which then sets the density of vortices via $v= \hbar/m (\nabla\phi)$ (see Ref. ). This is the correlation time  measured by Armitage[lvi] by terahertz spectroscopy (see Fig. 17); that group unfortunately interprets only the critical region as a spin liquid, ignoring the characteristic vortex liquid properties.

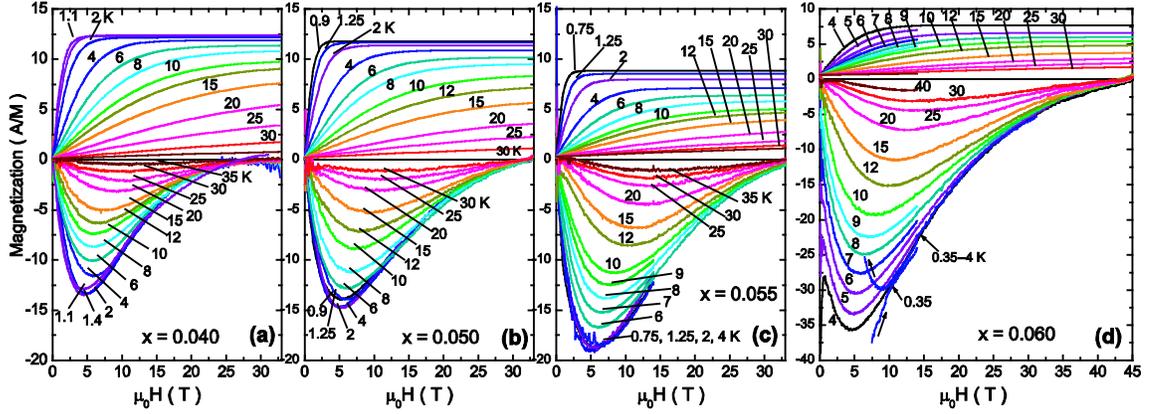

Figure 19 Planar diamagnetism (after subtraction of paramagnetic (vortex core?) signal) in very underdoped LASCO) [Ref. 47]

In Ref. 51, I derived what I called the "magic formula" for the actual shape of the Nernst or the equivalent diamagnetic response, a formula which is remarkably successful in describing the overall shape, though neither the experiments nor the theory are adequate to describe fine details, especially the precise behavior at low $B$.

The derivation is remarkably simple. It assumes that the pair current $\rho_s$ from each patch along the Fermi surface is additive and is decreased by the effects of temperature and magnetic field according to the standard formulas of BCS. The energy of a patch with gap $\Delta$, then, is
$$U = (B/\Phi_0)\rho_s(\Delta,T)\ln\frac{H_{c2}}{B}; \quad B < H_{c2}$$
$$U = 0, B > H_{c2}$$
where $H_{c2} = \Phi_0/\xi^2 = (c/2e\hbar v_F^2)\Delta^2 = K\Delta^2$
$\Phi_0$ is the flux quantum, and $\xi$ the coherence length $\hbar v_F/\Delta$.

To get a crude dependence on $T$ we approximate the usual formula by a step function:
$$\rho_s = 0, kT > \Delta$$
$$\rho_s = \rho_s(0), kT < \Delta$$

The *d*-wave gap then varies around the Fermi surface so that we have to integrate the energy over a gap distribution

$$P(\Delta)d\Delta = 1/\sqrt{(1-\Delta^2/\Delta_{max}^2)}d\Delta/\Delta_{max}$$

and the final formula for the Nernst effect is

$$\alpha_{xy} \propto B \int_{\max\{T,\sqrt{B/K}\}}^{\Delta_{max}} p(\Delta)d\Delta \ln(K\Delta^2/B)$$

This crude approximate formula is plotted as a function of *B* for various temperatures above $T_c$ in Fig. 17. I consider the resemblance to the general shape of data both for Nernst effect and for nonlinear diamagnetism, shown in accompanying Figs. 18 and 19, including its variation with *T*, to give no room for doubt as to the essential correctness of the vortex mechanism. It is a pity that more quantitative efforts have not been devoted to deriving the information that this formulation has made available, but unfortunately so much energy in the field of cuprates is wasted in unnecessary controversy that so far it has not been possible.

With these preliminary forays into the unique physical system constituting the vortex liquid my account must be closed. There are still many deep mysteries associated with these marvelous materials, but as far as I can see none of them will affect the foundational structure I outlined above.

# ACKNOWLEDGEMENTS

In a twenty-five year continuing effort there have been far too many collaborators and friends involved for me to hope to do all justice: I can here only pick out a few for special mention. The original RVB gang at Princeton: Baskaran, Z Zhou, Ted Hsu, Joe Wheatley, Shoudan Liang, Benoit Doucot, Sriram Shastry, Masao Ogata, later Steve Strong and David Clarke, along with some borrowed grad students like Stafford and Wen and colleagues like Ian Affleck, powered the first somewhat scattershot attack. The only colleague who has stayed with me essentially through the whole period has been Phuan Ong, and I owe him more than I can say; his and his students' experiments have been the most reliable guide through the whole mess. Various institutions have been particularly helpful at special times: IBM Research Laboratories, who helped fund the initial effort; Aspen Center for Physics, which provided an ideal milieu in which to achieve the necessary rethinking 1996-2001; CIAR, now CIFAR, where contacts, especially with Seamus Davis and with the UBC group, got me started on several of the recent advances discussed above. Other experimental groups who were generous with data and ideas were the Argonne ARPES group around Campuzano and Norman; Ali Yazdani's group; and particularly Kam Moler and John Kirtley who caused a vital turnaround. Finally, I owe special gratitude for the friendship and the reliable listening capacity of my "Besso", V Muthukumar; and, in the last stages, for the labor of my last student, Phil Casey.

[i] P W Anderson, Material Res Bull **8**, 153 (1973)
[ii] Walter Kohn, Phys Rev 133, A **171** (1964)
[iii] C Gros, T M Rice and R Joynt, Phys Rev **B** 36, 381 (1987)
[iv] G Baskaran, Z Zou, and P W Anderson, Solid State Comm **63**, 973 (1987). Several of our early papers contained this comment, which nonetheless is nowadays ascribed to a 1995 paper of Emery and Kivelson.
[v] S A Kivelson, D Rokhsar, J P Sethna, Phys Rev B **35**, 8865 (1987)
[vi] G Kotliar and J Liu, Phys Rev B38, 5132 (1988)
[vii] F C Zhang, C Gros, T M Rice , and Shiba, J Superc Sci & Tech **1**, 1 (1988)
[viii] P Anderson, Phys Rev Lett **96**, 017001 (2005)
[ix] see also B Edegger et al, Proc Natl Acad Sci **103**, 14298 (2006) for some references on this point.
[x] P Morel, P W Anderson, Phys Rev **123**, 1911 (1961)
[xi] K Krishana, J M Harris, and N P Ong, Phys Rev Lett **75**, 3529(1995)
[xii] P W Anderson, Physica B **318**, 28(2002)
[xiii] A Paramekanti, N Trivedi, M Randeria, Phys Rev Lett **87**, 217002 (2001)
[xiv] M Norman, J-C Campuzano, et al, Phys Rev B **54**, R9678 (1996)
[xv] W N Hardy et al, Phys Rev Lett **70**, 3999 (1993)
[xvi] Z X Shen et al, Phys Rev Lett **70**, 1553 (1993)
[xvii] D A Wollman, D J Van Harlingen, et al, Phys Rev Lett **71**, 2134 (1993)
[xviii] K A Moler, J Kirtley, et al, Science **279**, 1193 (1998)
[xix] P Monthoux and D Pines, Phys Rev B **47**, 6069 (1993) and further papers.
[xx] P W Anderson, P A Lee, M Randeria, T M Rice , N Trivedi, F C Zhang, J Phys Cond Mat R **754**, (2004)
[xxi] C Renner and O Fischer, Phys Rev B **51**, 9208 (1995)
[xxii] S H Pan, J C Davis, et al, Nature **403**, 746 (2000)
[xxiii] W Harrison, Phys Rev **123**, 85(1961)
[xxiv] P W Anderson, Concepts in Solids, Benjamin, N Y, (1963)
[xxv] B A Bernevig, R B Laughlin, D I Santiago, Phys Rev Lett **91**, 147003 (2003)
[xxvi] K Fujita, A Balatsky, et al, Phys Rev B **78**, 054510 (2008)
[xxvii] W Rantner and X G Wen, Phys Rev Lett **85**, 1722(2000)
[xxviii] M Tinkham, Phys Rev B **6**, 1747 (1972)
[xxix] P W Anderson and N P Ong, J Phys Chem Solids **67**, 1(2006)
[xxx] S C Davis, et al, Science 315, 1380 (2007) ; M Gomes, A Yazdani et al, Nature **447**, 569 (2007)
[xxxi] P W Anderson, Phys Rev Lett **104**, 176403 (2010)
[xxxii] G D Mahan, Phys Rev **163**, 612 (1967)
[xxxiii] P Nozieres and C deDominicis, Phys Rev **178**, 1097(1969)
[xxxiv] P W Anderson, Les Houches Lectures 1967, C Dewitt and R Balian, eds, Gordon and Breach (1968)
[xxxv] S W Tsai, A H Castro Neto, et al, J Phys Chem Solids **67**, 516 (2006)
[xxxvi] P W Anderson and G Yuval, J Phys C **3**, 2436 (1970)
[xxxvii] P A Casey et al, Nature Physics **4**, 208 (2008)

[xxxviii] P W Anderson, Nature Physics **2**, 626 (2006)
[xxxix] P A Casey & P W Anderson, Phys Rev B **80**, 094508(2009)
[xl] P W Anderson, Phys Rev Lett **18**, 1049 (1967)
[xli] N Hussey et al, Nature **415**, 814 (2003); Phys Rev B **76**, 104523 (2007); etc
[xlii] P A Casey, *thesis*, Princeton, (2010)
[xliii] In much of this discussion I am following K-Y Yang, T M Rice, F-C Zhang, Phys Rev B **73**, 174501 (2006) and merely fitting their ideas into the general "Hidden" scheme.
[xliv] A. Kanigel, U. Chatterjee, M. Randeria, M. R. Norman, S. Souma, M. Shi, Z. Z. Li, H. Raffy, and J. C. Campuzano, Phys Rev Lett **99**, 157001 (2007)
[xlv] S C Riggs, G S Boebinger et al, cond-mat /1008.1568 (2010)
[xlvi] Yayu Wang, Lu Li, N P Ong, Phys Rev B **73**, 024510 (2006)
[xlvii] Lu Li, N P Ong,, et al, Europhys Lett **72**, 451 (2005) (cond-mat/0507617); Lu Li et al, Nature Phys **3**, 311 (2007)
[xlviii] I Ussishkin,S L Sondhi, D A Huse, Phys Rev Lett **89**, 287001 (2002)
[xlix] G A Williams, Phys Rev Lett **61**, 1142 (1988)
[l] M J Salamon, et al, Phys Rev B **47**, 5520 (1993)
[li] P W Anderson, cond-mat/0603726 (2006); Nature Physics **3**, 160 (2007)
[lii] V Oganesyan, D L Huse, S L Sondhi, Phys Rev B **74**, 024425 (2006); S Mukherjee and D A Huse, Phys Rev B **72**, 064514 (2005)
[liii] R P Feynman, Prog Low Temp Phys **1**, 17-53 (1954) C J Gorter ed.
[liv] P W Anderson, cond-mat/0603726; Phys Rev Lett **100**, 215301 (2008)
[lv] S A Hartnoll et al, Phys Rev B **76**, 144502. (2009) This reference does not start from first principles, but hypothecates a quantum critical point dominating the phase diagram of the cuprates. It makes no attempt to confront the actual experimental anomalies, including the most striking, the existence of nonlinear diamagnetism over a wide range of the phase diagram. The physics, being based on a conformal field theory, has no possibility of resembling that of the actual substances, which are characterized by strong local interactions in a tight-binding band with an upper limit on the kinetic energy.
[lvi] L S Bilbro, H V Aguilar, G Loganov, I Bozovic, N P Armitage, submitted to Nature Physics (2010)